\documentclass[aps,prb,twocolumn,superscriptaddress]{revtex4-1}

\usepackage[utf8]{inputenc} 
\usepackage[english]{babel} 
\usepackage{graphicx} 
\usepackage{amssymb,amsmath} 
\usepackage[hidelinks]{hyperref} 
\usepackage{color}
\usepackage{siunitx} 
\usepackage{multirow}
\hypersetup{
	colorlinks,
	citecolor=blue,
	linkcolor=black,
	urlcolor=blue,
}

\begin{document}

\title{On the possibility to realize spin-orbit-induced correlated physics in iridium fluorides}

\author{M.~Rossi}
\affiliation{ESRF, The European Synchrotron, 71 Avenue des Martyrs, 38000 Grenoble, France}

\author{M.~Retegan}
\affiliation{ESRF, The European Synchrotron, 71 Avenue des Martyrs, 38000 Grenoble, France}

\author{C.~Giacobbe}
\affiliation{ESRF, The European Synchrotron, 71 Avenue des Martyrs, 38000 Grenoble, France}

\author{R.~Fumagalli}
\altaffiliation[Present address: ]{Dipartimento di Fisica, Politecnico di Milano, Piazza Leonardo da Vinci 32, 20133 Milano, Italy.}
\affiliation{ESRF, The European Synchrotron, 71 Avenue des Martyrs, 38000 Grenoble, France}

\author{A.~Efimenko}
\affiliation{ESRF, The European Synchrotron, 71 Avenue des Martyrs, 38000 Grenoble, France}

\author{T.~Kulka}
\affiliation{Institute of Theoretical Physics, Faculty of Physics, University of Warsaw, Pasteura 5, PL-02093 Warsaw, Poland}

\author{K.~Wohlfeld}
\affiliation{Institute of Theoretical Physics, Faculty of Physics, University of Warsaw, Pasteura 5, PL-02093 Warsaw, Poland}

\author{A.~I.~Gubanov}
\affiliation{Nikolaev Institute of Inorganic Chemistry, Siberian Branch of the Russian Academy of Sciences, Akademician Lavrentiev Prospekt 3, Novosibirsk 90, 630090, Russia}
\affiliation{Novosibirsk State University, Pirogova Street 2, Novosibirsk 90, 630090 Russia}

\author{M.~Moretti~Sala}
\email{marco.moretti@esrf.fr}
\affiliation{ESRF, The European Synchrotron, 71 Avenue des Martyrs, 38000 Grenoble, France}

\begin{abstract}

Recent theoretical predictions of ``unprecedented proximity'' of the electronic ground state of iridium fluorides to the SU(2) symmetric $j_{\mathrm{eff}}=1/2$ limit, relevant for superconductivity in iridates, motivated us to investigate their crystal and electronic structure. To this aim, we performed high-resolution x-ray powder diffraction, Ir L$_3$-edge resonant inelastic x-ray scattering, and quantum chemical calculations on Rb$_2$[IrF$_6$] and other iridium fluorides. Our results are consistent with the Mott insulating scenario predicted by Birol and Haule [Phys. Rev. Lett. 114, 096403 (2015)], but we observe a sizable deviation of the $j_{\mathrm{eff}}=1/2$ state from the SU(2) symmetric limit. Interactions beyond the first coordination shell of iridium are negligible, hence the iridium fluorides do not show any magnetic ordering down to at least 20 K. A larger spin-orbit coupling in iridium fluorides compared to oxides is ascribed to a reduction of the degree of covalency, with consequences on the possibility to realize spin-orbit-induced strongly correlated physics in iridium fluorides.

\end{abstract}

\maketitle

\section{Introduction}

The strong motivation behind the intense effort devoted to the investigation of iridium oxides (iridates) resides in the correlated nature of their physical properties. The identification of a spin-orbit-induced Mott insulating state in Sr$_2$IrO$_4$~\cite{Kim2008} triggered a number of theoretical and experimental studies aiming at isolating even more exotic phenomena, such as Kitaev-type magnetism~\cite{Jackeli2009,Takayama2015,Chun2015,Chaloupka2016} or Weyl semi-metallicity~\cite{Yang2010,Wan2011,Witczak-Krempa2012,Donnerer2016a}. For the specific case of Sr$_2$IrO$_4$, increasing experimental evidences of similarities with the high-temperature superconducting cuprates have been found in the structural, magnetic and electronic properties~\cite{Crawford1994,Kim2008,Kim2012}. It was therefore proposed that the low-energy physics of Sr$_2$IrO$_4$ could be described by a (pseudo)spin 1/2 particle in a one-band Hubbard model~\cite{Wang2011}, similarly to the cuprate antiferromagnetic (AFM) parent compounds, where the active orbital is branched off from the 5$d$--$t_{2g}$ states by virtue of strong spin-orbit coupling and it is usually termed the $j_{\mathrm{eff}}=1/2$ state~\cite{Kim2008}. Starting from the assumption that the one-orbital Hubbard model is a good approximation of the electronic structure of Sr$_2$IrO$_4$ and that high-temperature superconductivity in doped cuprates is described by the one band Hubbard model, unconventional superconductivity was said to be possible in doped iridates~\cite{Wang2011}. Superconductivity was theoretically predicted for both electron-~\cite{Watanabe2013,Yang2014} and hole-doped Sr$_2$IrO$_4$~\cite{Yang2014}. These results motivated a substantial experimental campaign to look for superconductivity in iridates, with encouraging results. It was shown that the AFM Mott insulating phase in Sr$_2$IrO$_4$ is destroyed upon electron doping and replaced by a paramagnetic phase with persistent magnetic excitations, strongly damped and displaying anisotropic softening~\cite{Gretarsson2016a}, in a way reminiscent of paramagnons in hole doped cuprates~\cite{LeTacon2011,LeTacon2013,Dean2013}. Angle-resolved photoemission spectroscopy measurements showed that electron doped Sr$_2$IrO$_4$ displays typical features of superconducting cuprates, such as Fermi arcs~\cite{Kim2014a,DeLaTorre2015} and a $d$-wave gap~\cite{Kim2016} in the intermediate and low temperature phases, respectively. Despite exciting theoretical predictions and promising experimental findings, however, superconductivity has not been observed yet in doped Sr$_2$IrO$_4$ or any other iridium oxide. 

The theoretical finding of spin-orbit-induced correlated physics in a novel class of materials exhibiting an ``unprecedented proximity'' to the ideal SU(2) limit is therefore extremely welcome~\cite{Birol2015}. Indeed, recently Birol and Haule~\cite{Birol2015} proposed the exciting idea that spin-orbit-induced correlated physics can be found in a novel class of materials, namely rhodium and iridium fluorides. A first indication that this may be the case comes from Pedersen \emph{et al.}~\cite{Pedersen2016} who showed that the magnetism of ideal model system molecular iridium fluorides is consistent with the $j_{\mathrm{eff}}=1/2$ scenario and that they can be used as building-blocks to synthesize electronic and magnetic quantum materials~\cite{Pedersen2016}, such as those proposed by Birol and Haule. Rb$_2$[IrF$_6$] is particularly appealing because it is said to host a $j_{\mathrm{eff}}=1/2$ ground state with ``unprecedented proximity'' to the SU(2) symmetric limit, with possible implications for superconductivity in iridates~\cite{Birol2015}. 

The main motivation behind our study is to understand differences and analogies between the physics of iridium fluorides and oxides. To this aim we investigate the crystal and electronic structure of several iridium fluorides (Rb$_2$[IrF$_6$], Na$_2$[IrF$_6$], K$_2$[IrF$_6$], Cs$_2$[IrF$_6$], and Ba[IrF$_6$]) by means of high-resolution x-ray powder diffraction (XRPD), resonant inelastic x-ray scattering (RIXS), and quantum chemical calculations. Our results are consistent with the predictions of a wide gap $j_{\mathrm{eff}}=1/2$ Mott insulator retaining a paramagnetic state~\cite{Birol2015} down to 20 K. Indeed, we find that the low-energy electronic structure of these systems is mostly dictated by the local coordination of the IrF$_6$ octahedra, with no evidence of interactions beyond the first coordination shell of iridium. We observe nevertheless a sizable deviation of the $j_{\mathrm{eff}}=1/2$ state from the SU(2) symmetric limit suggesting that the distortions in the electronic structure due to a non-cubic environment are larger than predicted. Our experimental results are supported by quantum chemical calculations.

\section{Experimental details}

High-resolution XRPD measurements were performed at beamline ID22 of the European Synchrotron Radiation Facility (ESRF, France). The incoming x-rays were monochromated to $\lambda = 0.3999$ \AA~by a Si(111) double-crystal monochromator. The x-rays diffracted by the sample were collimated by 9 Si(111) analyzers and collected by a Cyberstar scintillation detector. 

Iridium L$_3$-edge RIXS spectra were measured at the inelastic x-ray scattering beamline ID20 of the ESRF. ID20 is particularly suited for RIXS experiments due to its energy resolution capabilities. This is as good as 15 meV at 11.2165 keV when a Si(844) back-scattering channel-cut is used to monochromate the incident photon beam. The spectrometer is based on a single Si(844) diced-crystal analyzer ($R = 1$ m) in Rowland scattering geometry and equipped with a two-dimensional Maxipix detector~\cite{Ponchut2011}. The overall energy resolution was set to 35 meV for this experiment~\cite{MorettiSala2013}.

Samples were grown in the Nikolaev Institute of Inorganic Chemistry (Novosibirsk, Russia). Na$_2$[IrF$_6$] was prepared following the method described in Ref.~\onlinecite{Pedersen2016}. After dissolving 1.009 g of Na$_2$[IrF$_6$] in 10 ml of H$_2$O, 10 ml of cation-resin H$^+$ were added to the solution. After 30 min of mixing and filtering, H$_2$[IrF$_6$] was obtained. K$_2$[IrF$_6$], Rb$_2$[IrF$_6$], Cs$_2$[IrF$_6$] and Ba[IrF$_6$] were prepared by filtering the solutions obtained after reaction of stoichiometric quantities of H$_2$[IrF$_6$] and KF, RbF, CsF and BaCO$_3$, respectively. Single crystals of Rb$_2$[IrF$_6$] were grown by slow counter diffusion of Na$_2$[IrF$_6$] (0.3 M) and RbF (20 M) solutions in 1\% agar gel. 

\section{Computational method}

Quantum chemical calculations were performed using the ORCA software package~\cite{Neese2012}. State-averaged complete active space self-consistent field (CASSCF) and $N$-electron valence perturbation theory (NEVPT2)~\cite{Angeli2001} calculations were used to determine the energy of the excited states, and the spin-orbit-coupling constant~\cite{Atanasov2011}. The active space included five electrons distributed over the five \textit{d}-orbitals. The spin-orbit coupling was treated a posteriori using the quasi-degenerate perturbation theory~\cite{Ganyushin2006} and the mean-field approximation of the Breit-Pauli spin-orbit coupling operator~\cite{Neese2005}. Scalar relativistic effects were included using the zero-order regular approximation (ZORA)~\cite{vanWullen1998}. Polarized triple-$\zeta$ basis sets were used for all elements~\cite{Weigend2005,Pantazis2008}. All computations were done using the embedded cluster approach in order to account for the environment~\cite{Staemmler2005}. Models constructed starting from the XRPD structure consisted of a central IrF$_6$ octahedron (the quantum cluster (QC)) surrounded by point charges (PC). A boundary region (BR) containing repulsive capped effective core potentials was introduced to avoid electron flow from the central subunit towards the point charges.

\section{Results and discussion}

\begin{figure}
\includegraphics[width=\columnwidth]{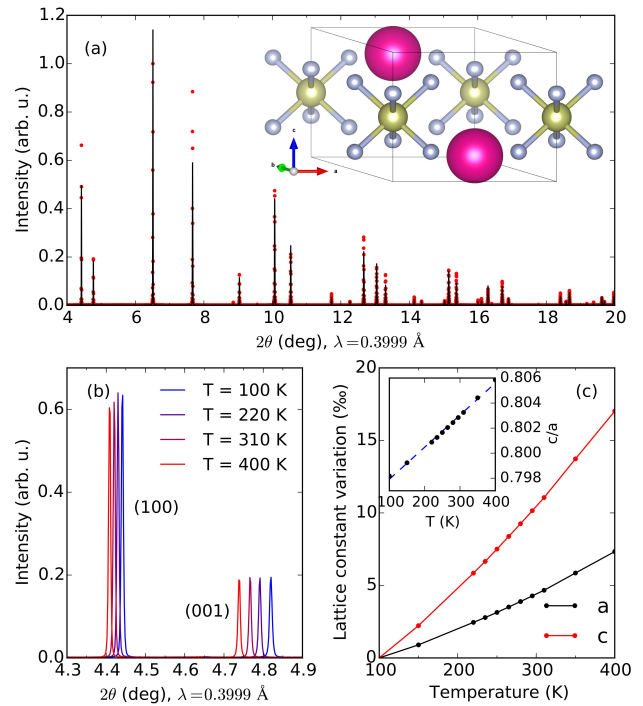}
\caption{\label{fig:Rb_powder_diffraction} (a) High-resolution XRPD measurements on Rb$_2$[IrF$_6$] performed at $\lambda = 0.3999$ \AA~and T = 295 K confirm that the space group of this compound is $P\bar{3}m1$ (164)\cite{Smolentsev2007}. Red dots are the data, the black solid line displays the Rietveld refinement. The corresponding crystal structure is shown in the inset. (b) The smooth angular variation of the (100) and (001) reflections with temperature reveals that the $a$ and $c$ lattice constants change differently with temperature and that no phase transition is observed in the range between 100 K and 400 K. (c) Temperature dependence of the lattice constants and their ratio.}
\end{figure}

\begin{table*}
\caption{\label{tab:Comparison_300K} Space group, site symmetry, Ir-F distance, F-Ir-F angle, trigonal angle ($\beta$), and octahedral distortion at T = 295 K for A$_2$[IrF$_6$] (A = Na, K, Rb, Cs) and Ba[IrF$_6$].}
\begin{ruledtabular}
\begin{tabular}{lccccccc}
Compound & Space group & Site symmetry & Ir-F (\AA) & F-Ir-F ($^\circ$) & $\beta$ ($^\circ$) & $(\beta-\beta_0)/\beta_0$ (\%)  \\ 
\hline
Na$_2$[IrF$_6$] & $P321$ & 32./3.. & 1.939/1.948 &  & 56.48/55.70 & 3.2/1.75  \\ 
K$_2$[IrF$_6$] & $P\bar{3}m1$ & $\bar{3}m.$ & 1.940 & 86.6/93.4 & 57.19 & 4.5  \\ 
Rb$_2$[IrF$_6$] & $P\bar{3}m1$ & $\bar{3}m.$ & 1.975 & 87.0/93.0 & 56.89 & 3.9  \\ 
Cs$_2$[IrF$_6$] & $P\bar{3}m1$ & $\bar{3}m.$ & 1.941 & 87.5/92.5 & 56.52 & 3.3  \\ 
Ba[IrF$_6$] & $R\bar{3}$ & $\bar{3}.$ & 1.937 & 85.9/94.1 & 51.89 & -5.2  \\ 
\end{tabular}
\end{ruledtabular}
\end{table*}

Figure~\ref{fig:Rb_powder_diffraction}(a) shows the high-resolution XRPD pattern of Rb$_2$[IrF$_6$] at $T = 295$ K. The black solid line corresponds to the Rietveld refinement of the experimental data (red dots). We find that Rb$_2$[IrF$_6$] belongs to the space group $P\bar{3}m1$ with lattice parameters $a = b = 5.9777(0)$~\AA~and $c = 4.7986(5)$~\AA~at 295 K. Crystallite size broadening between 4.8 and \SI{5.2}{\micro\meter} has been estimated using an instrumental peak shape function implemented in Topas 5 \cite{Coelho2007} and convolving sample size term on top \cite{Balzar2004}. As it can be seen in the inset, the crystal structure of Rb$_2$[IrF$_6$] consists of isolated IrF$_6$ octahedra. The Ir-F bond lengths are all the same and equal to 1.975~\AA, while the F-Ir-F bond angles are 87$^\circ$ and 93$^\circ$. As a result, the octahedra are slightly compressed along the crystallographic $c$ axis, thus inducing a trigonal distortion of 3.9\%, defined as $(\beta-\beta_0)/\beta_0$, where $\beta_0\approx54.74^\circ$ and $\beta$ is the angle between the Ir-F bond and the trigonal axis~\cite{Avram2013}. A similar analysis has been carried out for all the compounds. The results of the crystal structure refinement, summarized in Table~\ref{tab:Comparison_300K}, are in agreement with existing literature~\cite{Smolentsev2007,Smolentsev2007a,Fitz2002}, except for Ba[IrF$_6$], for which we converged to the $R\bar{3}$ space group \cite{Smolentsev2007b}. The common feature to all systems is the presence of isolated IrF$_6$ units with comparable distortions of the octahedral cage. In the case of Rb$_2$[IrF$_6$], XRPD measurements were performed at several temperatures in the range between 100 and 400 K. Fig.~\ref{fig:Rb_powder_diffraction}(b) shows the temperature dependence of the diffraction peaks associated to the (100) and the (001) reflections. Their continuous variation can be directly associated to a smooth change of the $a$ and $c$ lattice parameters, as reported in Fig.~\ref{fig:Rb_powder_diffraction}(c), and suggests that no structural phase transition occurs in the investigated temperature range.

\begin{figure}
\includegraphics[width=\columnwidth]{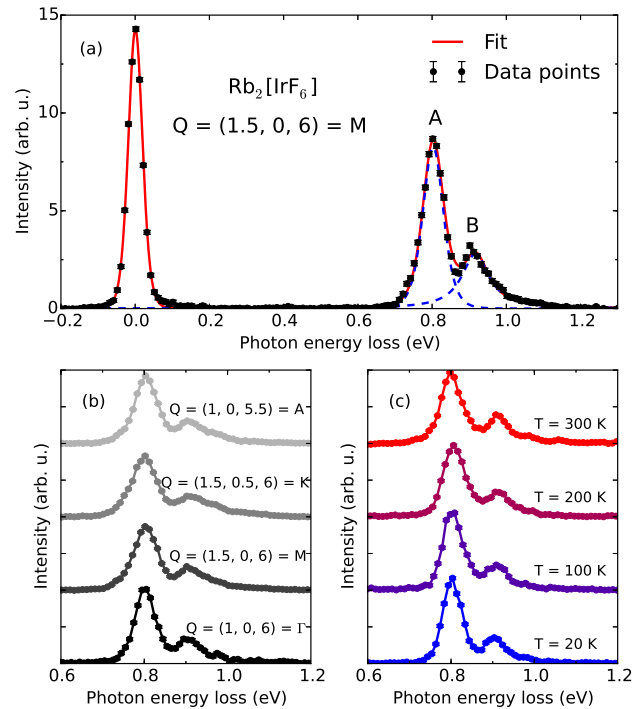}
\caption{\label{fig:Rb_RIXS} (a) Ir L$_3$-edge RIXS spectrum of Rb$_2$[IrF$_6$] measured at $T = 20$ K. Black dots are the background-subtracted experimental points, the red solid line is the total fit, and the dotted blue curves are the fit to the $j_{\mathrm{eff}}=3/2$ excitations. (b) Momentum transfer dependence of the $j_{\mathrm{eff}}=3/2$ excitations at the high-symmetry points of the Brillouin zone. (c) Temperature dependence of the $j_{\mathrm{eff}}=3/2$ excitations between 20 K and 300 K, measured on powder Rb$_2$[IrF$_6$].}
\end{figure}

\begin{figure}
\includegraphics[width=\columnwidth]{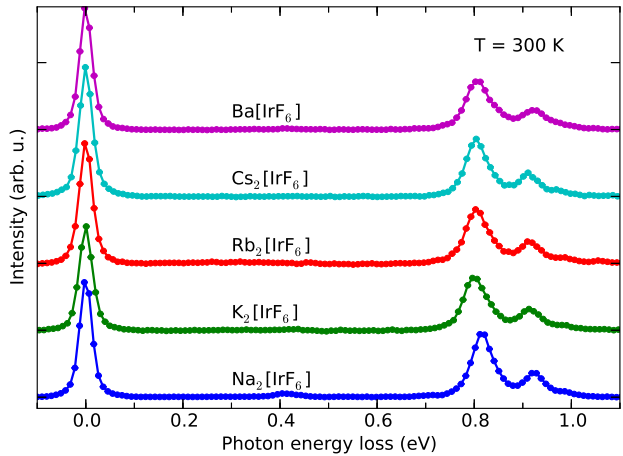}
\caption{\label{fig:Comparison_300K} Ir L$_3$-edge RIXS spectra of A$_2$[IrF$_6$] (A = Na, K, Rb, Cs) and Ba[IrF$_6$]. All spectra clearly show the presence of two $j_\mathrm{eff} = 3/2$ excitations.}
\end{figure}

After characterizing the samples from a structural point of view, we now turn to the investigation of their electronic structure. Figure~\ref{fig:Rb_RIXS}(a) shows a representative Ir L$_3$-edge RIXS spectrum of Rb$_2$[IrF$_6$] single crystal measured at momentum transfer $\mathbf{Q} = (1.5, 0, 6)$ r.l.u.~and $T = 20$ K. The incident photon energy was fixed at 11.2165 keV, i.e. $\sim 3$ eV below the main absorption line, where intra-$t_{2g}$ excitations are enhanced~\cite{Ishii2011,MorettiSala2014b,Lefrancois2016}. The black dots in Fig.~\ref{fig:Rb_RIXS}(a) correspond to the background-subtracted data points, while the red solid line is the fit to the data. We highlight the absence of features below 0.7 eV. At higher energy losses, two features (A and B) are clearly distinguished. They are fitted to two Pearson VII functions~\cite{Michette2000} (blue dashed lines in Fig.~\ref{fig:Rb_RIXS}(a)) and their energy positions are $805\pm 0.4$ meV (A) and $915\pm 1.3$ meV (B). Considering the resonance behavior of the two features, we ascribe them to transitions from the $j_{\mathrm{eff}}=1/2$ to the  $j_{\mathrm{eff}}=3/2$ states, in line with previous RIXS studies of iridium oxides~\cite{Kim2012,Gretarsson2013,Kim2014, MorettiSala2014b,Lefrancois2016}. In addition, the two features do not show any detectable momentum or temperature dependence within the experimental uncertainties, as shown in Figs.~\ref{fig:Rb_RIXS}(b) and (c), suggesting that the IrF$_6$ octahedra behave as isolated units. Our findings support the scenario of a strong insulating character of iridium fluorides, with narrow bands and no tendency to develop long-range magnetic order, in line with theoretical predictions~\cite{Birol2015}. Similar measurements were carried out for all samples in powder form. Figure~\ref{fig:Comparison_300K} shows a stack of the corresponding RIXS spectra. Interestingly, the overall shape is very similar and closely resembles the RIXS spectrum of Fig.~\ref{fig:Rb_RIXS}(a). However, before discussing the small but meaningful differences between the different compounds, we notice that they all show a large splitting of the $j_\mathrm{eff} = 3/2$ states. This is indicative of a sizable lifting of the $5d$--$t_{2g}$ states degeneracy~\cite{MorettiSala2014a} and contrasts with the prediction of an isotropic electronic state close to the SU(2) limit for Rb$_2$[IrF$_6$]~\cite{Birol2015}.

\begin{table}
\caption{\label{tab:exp_vs_th} Energy position and splitting of the A and B features, and calculation of spin-orbit-coupling constant ($\zeta$) and trigonal distortion ($\Delta$) to the cubic crystal field for the different compounds. All values are given in meV.}
\begin{ruledtabular}
\begin{tabular}{lcccccccc}
Compound & $E_A^\mathrm{exp}$ & $E_B^\mathrm{exp}$ & $\Delta_{BA}^\mathrm{exp}$ & $\Delta_{BA}^\mathrm{calc}$ & $\zeta$ & $\Delta$\\ 
\hline
\multirow{2}{*}{Na$_2$[IrF$_6$]}   & \multirow{2}{*}{$816\pm 0.3$} & \multirow{2}{*}{$923\pm 0.7$} & \multirow{2}{*}{$107\pm 0.8$} & 19 & \multirow{2}{*}{574} & \multirow{2}{*}{-152}\\
                                   &                      &                      &                      & 50 &                      & \\
K$_2$[IrF$_6$]                     & $802\pm 0.3$         & $914\pm 1.0$         & $112\pm 1.1$         & 53 & 566                  & -159\\
Rb$_2$[IrF$_6$]                    & $805\pm 0.4$         & $915\pm 1.3$         & $110\pm 1.4$         & 44 & 567                  & -156\\
Cs$_2$[IrF$_6$]                    & $804\pm 0.5$         & $911\pm 1.5$         & $107\pm 1.6$         & 33 & 566                  & -152\\
Ba[IrF$_6$]                        & $807\pm 0.5$         & $921\pm 1.7$         & $114\pm 1.8$		    & 71 & 567                  &  186\\
\end{tabular}
\end{ruledtabular}
\end{table}

\begin{figure}
\includegraphics[width=\columnwidth]{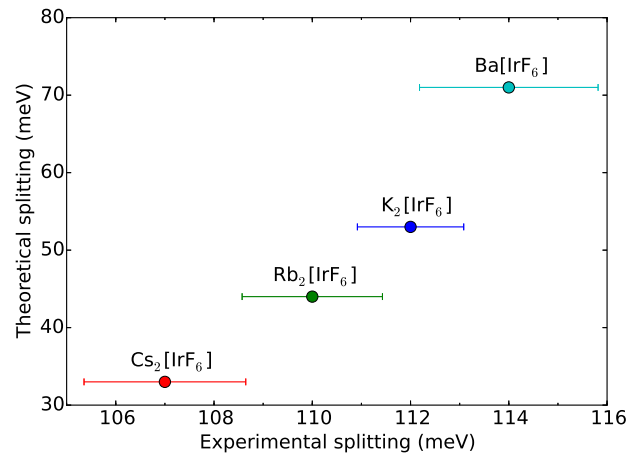}
\caption{\label{fig:exp_vs_th} Splitting of the $j_\mathrm{eff} = 3/2$ excitations, as results from experimental data and quantum chemical calculations.}
\end{figure}

To get a better insight into the electronic structure of iridium fluorides, we have performed quantum chemical calculations using the embedded cluster approach. The calculated splittings of the $j_\mathrm{eff} = 3/2$ excited states are compared to the experimental values in Table~\ref{tab:exp_vs_th}. Although the agreement with the experiment is not quantitative, the calculations reproduce the trend among the different compounds. As can be seen in Fig.~\ref{fig:exp_vs_th}, the calculated and experimental splittings are found to correlate very nicely (we exclude Na$_2$[IrF$_6$] from this analysis because there are two inequivalent iridium sites in this compound). Overall, the agreement between experiments and quantum chemical calculations suggests that there is little or no influence of the alkali (earth) metal on the low-energy electronic structure of iridium fluorides, but rather that it is solely dictated by the local coordination of the IrF$_6$ octahedra. This is mainly the consequence of the fact that iridium fluorides are composed by disconnected IrF$_6$ units. We note that while chemical substitution is not effective at modifying the low-energy electronic structure of iridium fluorides, physical pressure may be.

In order to compare iridium fluorides and iridates, we have expressed the experimental results presented above in terms of single-ion model parameters, such as the effective trigonal distortion $\Delta$ of the cubic crystal field, and the spin-orbit-coupling constant $\zeta$, often used in the literature of iridates~\cite{Jackeli2009,Ament2011,Liu2012,Boseggia2013,Perkins2014,MorettiSala2014a,MorettiSala2014b,MorettiSala2014c,MorettiSala2014d,Lefrancois2016,DiMatteo2016}. By constraining the energies of the $j_\mathrm{eff} = 3/2$ excited states as calculated from the single-ion model to the energy positions of the features A and B, and by taking into account the sign of the octahedral distortion as determined by XRPD, we estimated the values of $\Delta$ and $\zeta$ for our systems. As reported in Table~\ref{tab:exp_vs_th}, estimates for $|\Delta|$ vary between 0.15 and 0.19 eV, while $\zeta\approx 0.57$ eV for all the iridium fluorides studied here. The latter is 10--40\% larger than in iridium oxides, where it ranges between 0.38~\cite{Kim2014} and 0.52 eV~\cite{MorettiSala2014b}. Ligand-field parameters, spin-orbit-coupling constant, and interelectronic repulsion terms have also been calculated by fitting the full configuration interaction matrix elements obtained from the CASSCF/NEVPT2 calculation to the matrix elements of a model Hamiltonian containing those interactions~\cite{Atanasov2011}. In agreement with the experimentally fitted values discussed above, we calculate $\zeta\approx 0.55$ eV for all iridium fluorides, and $\zeta\approx 0.52$ eV for Sr$_2$IrO$_4$ using the same theoretical approach. In order to rule out the possibility that the reduction of spin-orbit coupling in iridium oxides compared to fluorides arises from differences in their crystal structure, we calculated the spin-orbit-coupling constant for an IrF$_6$ octahedron where the F ions have been placed at the positions of the O ions in Sr$_2$IrO$_4$. We obtain $\zeta\approx 0.55$ eV, suggesting that the nature of the coordinating ion, rather than the crystal structure, determines the differences in the spin-orbit-coupling constant. The larger reduction of $\zeta$ compared to the free ion value (relativistic nephelauxetic effect~\cite{Jorgensen1963}) in iridium fluorides than in oxides reflects the smaller degree of covalency in the chemical bonds of the former. This is of particular interest because covalency in Sr$_2$IrO$_4$ is thought to be responsible for strong orbital anisotropies, in view of the increased spatial extent of the 5$d$--$t_{2g}$ orbitals reaching the nearest neighbor iridium atoms and beyond~\cite{Agrestini2017}. Iridium fluorides might therefore be the ideal playground for studying spin-orbit-induced correlated physics because correlation effects might be enhanced by the more localized nature of the electronic states, whereas long-range anisotropies, which contribute to deviate the $j_\mathrm{eff} = 1/2$ from the SU(2) symmetric limit, are strongly suppressed. We note that a reduction of covalency would lower the ratio of the energy scales of the magnetic over the charge fluctuations. The latter effect might become important may the iridium fluorides be electron- or hole-doped.

As a final remark, we would like to discuss oxides and fluorides in relation to the similarities between cuprates and iridates. We start by considering La$_2$CuO$_4$ and K$_2$CuF$_4$. Although they share the same K$_2$[NiF$_4$]-type crystal structure and are insulating, their magnetic properties are very distinct: La$_2$CuO$_4$ is an AFM insulator~\cite{Vaknin1987,Chen1991}, while K$_2$CuF$_4$ has a ferromagnetic (FM) ground state~\cite{Ito1976,Kugel1982}. Indeed, in La$_2$CuO$_4$ a strong tetragonal crystal field splits the 3$d$--$e_g$ states and stabilizes the $x^2-y^2$ orbital, which gives rise to ferro-orbital ordering and ultimately to AFM coupling on a straight (180$^\circ$) bond. On the contrary, in the undistorted K$_2$CuF$_4$ the degeneracy of the 3$d$--$e_g$ states is essentially preserved and the Cu$^{2+}$ ion is Jahn-Teller active. The so-called cooperative Jahn-Teller effect sets in and $x^2-z^2$/$y^2-z^2$ alternating orbital ordering is stabilized leading to FM long-range order in the ground state~\cite{Kugel1982}. When moving to Sr$_2$IrO$_4$, much of the physics of La$_2$CuO$_4$ is retained, namely the Mott insulating AFM state with dominant Heisenberg-like interactions~\cite{Kim2008,Kim2012}. Effectively, the only active orbital is the $j_{\mathrm{eff}}=1/2$, which is branched off from the 5$d$--$t_{2g}$ by virtue of the strong spin-orbit coupling. One important consequence of such strong spin-orbit coupling is that, no matter how undistorted the system is, the Jahn-Teller mechanism is not supported in the Ir$^{4+}$ compounds~\cite{Plotnikova2016}. We therefore speculate that while the lack of magnetism in the studied iridium fluorides can basically be attributed to the isolation of the IrF$_6$ units, the ground state of a hypothetical iridium fluoride with an ideal K$_2$[NiF$_4$]-type crystal structure would probably never support FM order (unlike copper fluorides). Instead, it might even be closer to the Heisenberg AFM state found in copper oxides than iridium oxides. In this respect, the parallelism between copper oxides/fluorides and iridium oxides/fluorides is broken.

\section{Conclusions}

We investigated the crystal and electronic structure of Na$_2$[IrF$_6$], K$_2$[IrF$_6$], Rb$_2$[IrF$_6$], Cs$_2$[IrF$_6$] and Ba[IrF$_6$] by means of high-resolution XRPD, Ir L$_3$-edge RIXS, and quantum chemical calculations. Our results support the theoretical predictions that Rb$_2$[IrF$_6$] is characterized by a $j_{\mathrm{eff}}=1/2$ electronic ground state~\cite{Birol2015}. The absence of low-energy features, as well as momentum and temperature dependence in the RIXS spectra of Rb$_2$IrF$_6$ single crystal suggests that interactions beyond the first coordination shell of iridium ions are negligible, thus precluding long-range magnetic order down to at least 20 K. However, the splitting of the $j_\mathrm{eff} = 3/2$ excited states is indicative of a deviation from the SU(2) symmetric limit. Consistently, quantum chemical calculations on a single IrF$_6$ cluster reproduce the experimental trend observed among the various compounds and elucidate that the low-energy electronic structure of iridium fluorides is ascribed to characteristic local distortions of the IrF$_6$ cage with no significant influence from neighboring ions.

We also report an increase of the spin-orbit coupling in iridium fluorides as compared to iridium oxides. This finding is corroborated by quantum chemical calculations and suggests that the larger electronegativity of fluorine compared to oxygen reduces the degree of covalency in the system. This has important consequences: i) the spatial extension of 5$d$--$t_{2g}$ orbitals is reduced and correlation effects might indeed be enhanced; ii) long-range anisotropies are mitigated and an isotropic $j_{\mathrm{eff}}=1/2$ ground state is more likely to stabilize. If synthesized in crystal structures with connected IrF$_6$ units, such hypothetical iridium fluorides might indeed support features characteristic of spin-orbit-induced strongly correlated physics and might even more closely resemble the low-energy physics found in copper oxides than it is the case of iridium oxides.

\begin{acknowledgments}

We acknowledge the European Synchrotron Radiation Facility (ESRF, France) for providing beamtime. T.~K. thanks the ESRF for kind hospitality. K.~W. acknowledges support by Narodowe Centrum Nauki (NCN, National Science Centre) under Project No. 2012/04/A/ST3/00331 and Project No. 2016/22/E/ST3/00560. The authors are grateful to C.~Henriquet and B.~Detlefs for technical assistance during the experiment, and to M.~Krisch for critical reading of the manuscript.

\end{acknowledgments}

\bibliography{library}

\begin{thebibliography}{62}%
\makeatletter
\providecommand \@ifxundefined [1]{%
 \@ifx{#1\undefined}
}%
\providecommand \@ifnum [1]{%
 \ifnum #1\expandafter \@firstoftwo
 \else \expandafter \@secondoftwo
 \fi
}%
\providecommand \@ifx [1]{%
 \ifx #1\expandafter \@firstoftwo
 \else \expandafter \@secondoftwo
 \fi
}%
\providecommand \natexlab [1]{#1}%
\providecommand \enquote  [1]{``#1''}%
\providecommand \bibnamefont  [1]{#1}%
\providecommand \bibfnamefont [1]{#1}%
\providecommand \citenamefont [1]{#1}%
\providecommand \href@noop [0]{\@secondoftwo}%
\providecommand \href [0]{\begingroup \@sanitize@url \@href}%
\providecommand \@href[1]{\@@startlink{#1}\@@href}%
\providecommand \@@href[1]{\endgroup#1\@@endlink}%
\providecommand \@sanitize@url [0]{\catcode `\\12\catcode `\$12\catcode
  `\&12\catcode `\#12\catcode `\^12\catcode `\_12\catcode `\%12\relax}%
\providecommand \@@startlink[1]{}%
\providecommand \@@endlink[0]{}%
\providecommand \url  [0]{\begingroup\@sanitize@url \@url }%
\providecommand \@url [1]{\endgroup\@href {#1}{\urlprefix }}%
\providecommand \urlprefix  [0]{URL }%
\providecommand \Eprint [0]{\href }%
\providecommand \doibase [0]{http://dx.doi.org/}%
\providecommand \selectlanguage [0]{\@gobble}%
\providecommand \bibinfo  [0]{\@secondoftwo}%
\providecommand \bibfield  [0]{\@secondoftwo}%
\providecommand \translation [1]{[#1]}%
\providecommand \BibitemOpen [0]{}%
\providecommand \bibitemStop [0]{}%
\providecommand \bibitemNoStop [0]{.\EOS\space}%
\providecommand \EOS [0]{\spacefactor3000\relax}%
\providecommand \BibitemShut  [1]{\csname bibitem#1\endcsname}%
\let\auto@bib@innerbib\@empty
\bibitem [{\citenamefont {Kim}\ \emph {et~al.}(2008)\citenamefont {Kim},
  \citenamefont {Jin}, \citenamefont {Moon}, \citenamefont {Kim}, \citenamefont
  {Park}, \citenamefont {Leem}, \citenamefont {Yu}, \citenamefont {Noh},
  \citenamefont {Kim}, \citenamefont {Oh}, \citenamefont {Park}, \citenamefont
  {Durairaj}, \citenamefont {Cao},\ and\ \citenamefont {Rotenberg}}]{Kim2008}%
  \BibitemOpen
  \bibfield  {author} {\bibinfo {author} {\bibfnamefont {B.~J.}\ \bibnamefont
  {Kim}}, \bibinfo {author} {\bibfnamefont {H.}~\bibnamefont {Jin}}, \bibinfo
  {author} {\bibfnamefont {S.~J.}\ \bibnamefont {Moon}}, \bibinfo {author}
  {\bibfnamefont {J.-Y.}\ \bibnamefont {Kim}}, \bibinfo {author} {\bibfnamefont
  {B.-G.}\ \bibnamefont {Park}}, \bibinfo {author} {\bibfnamefont {C.~S.}\
  \bibnamefont {Leem}}, \bibinfo {author} {\bibfnamefont {J.}~\bibnamefont
  {Yu}}, \bibinfo {author} {\bibfnamefont {T.~W.}\ \bibnamefont {Noh}},
  \bibinfo {author} {\bibfnamefont {C.}~\bibnamefont {Kim}}, \bibinfo {author}
  {\bibfnamefont {S.-J.}\ \bibnamefont {Oh}}, \bibinfo {author} {\bibfnamefont
  {J.-H.}\ \bibnamefont {Park}}, \bibinfo {author} {\bibfnamefont
  {V.}~\bibnamefont {Durairaj}}, \bibinfo {author} {\bibfnamefont
  {G.}~\bibnamefont {Cao}}, \ and\ \bibinfo {author} {\bibfnamefont
  {E.}~\bibnamefont {Rotenberg}},\ }\href {\doibase
  10.1103/PhysRevLett.101.076402} {\bibfield  {journal} {\bibinfo  {journal}
  {Phys. Rev. Lett.}\ }\textbf {\bibinfo {volume} {101}},\ \bibinfo {pages}
  {076402} (\bibinfo {year} {2008})},\ \Eprint {http://arxiv.org/abs/0803.2927}
  {arXiv:0803.2927} \BibitemShut {NoStop}%
\bibitem [{\citenamefont {Jackeli}\ and\ \citenamefont
  {Khaliullin}(2009)}]{Jackeli2009}%
  \BibitemOpen
  \bibfield  {author} {\bibinfo {author} {\bibfnamefont {G.}~\bibnamefont
  {Jackeli}}\ and\ \bibinfo {author} {\bibfnamefont {G.}~\bibnamefont
  {Khaliullin}},\ }\href {\doibase 10.1103/PhysRevLett.102.017205} {\bibfield
  {journal} {\bibinfo  {journal} {Phys. Rev. Lett.}\ }\textbf {\bibinfo
  {volume} {102}},\ \bibinfo {pages} {2} (\bibinfo {year} {2009})},\ \Eprint
  {http://arxiv.org/abs/0809.4658} {arXiv:0809.4658} \BibitemShut {NoStop}%
\bibitem [{\citenamefont {Takayama}\ \emph {et~al.}(2015)\citenamefont
  {Takayama}, \citenamefont {Kato}, \citenamefont {Dinnebier}, \citenamefont
  {Nuss}, \citenamefont {Kono}, \citenamefont {Veiga}, \citenamefont {Fabbris},
  \citenamefont {Haskel},\ and\ \citenamefont {Takagi}}]{Takayama2015}%
  \BibitemOpen
  \bibfield  {author} {\bibinfo {author} {\bibfnamefont {T.}~\bibnamefont
  {Takayama}}, \bibinfo {author} {\bibfnamefont {A.}~\bibnamefont {Kato}},
  \bibinfo {author} {\bibfnamefont {R.}~\bibnamefont {Dinnebier}}, \bibinfo
  {author} {\bibfnamefont {J.}~\bibnamefont {Nuss}}, \bibinfo {author}
  {\bibfnamefont {H.}~\bibnamefont {Kono}}, \bibinfo {author} {\bibfnamefont
  {L.~S.~I.}\ \bibnamefont {Veiga}}, \bibinfo {author} {\bibfnamefont
  {G.}~\bibnamefont {Fabbris}}, \bibinfo {author} {\bibfnamefont
  {D.}~\bibnamefont {Haskel}}, \ and\ \bibinfo {author} {\bibfnamefont
  {H.}~\bibnamefont {Takagi}},\ }\href {\doibase
  10.1103/PhysRevLett.114.077202} {\bibfield  {journal} {\bibinfo  {journal}
  {Phys. Rev. Lett.}\ }\textbf {\bibinfo {volume} {114}} (\bibinfo {year}
  {2015}),\ 10.1103/PhysRevLett.114.077202},\ \Eprint
  {http://arxiv.org/abs/1403.3296} {arXiv:1403.3296} \BibitemShut {NoStop}%
\bibitem [{\citenamefont {{Hwan Chun}}\ \emph {et~al.}(2015)\citenamefont
  {{Hwan Chun}}, \citenamefont {Kim}, \citenamefont {Kim}, \citenamefont
  {Zheng}, \citenamefont {Stoumpos}, \citenamefont {Malliakas}, \citenamefont
  {Mitchell}, \citenamefont {Mehlawat}, \citenamefont {Singh}, \citenamefont
  {Choi}, \citenamefont {Gog}, \citenamefont {Al-Zein}, \citenamefont {Sala},
  \citenamefont {Krisch}, \citenamefont {Chaloupka}, \citenamefont {Jackeli},
  \citenamefont {Khaliullin},\ and\ \citenamefont {Kim}}]{Chun2015}%
  \BibitemOpen
  \bibfield  {author} {\bibinfo {author} {\bibfnamefont {S.}~\bibnamefont
  {{Hwan Chun}}}, \bibinfo {author} {\bibfnamefont {J.-W.}\ \bibnamefont
  {Kim}}, \bibinfo {author} {\bibfnamefont {J.}~\bibnamefont {Kim}}, \bibinfo
  {author} {\bibfnamefont {H.}~\bibnamefont {Zheng}}, \bibinfo {author}
  {\bibfnamefont {C.~C.}\ \bibnamefont {Stoumpos}}, \bibinfo {author}
  {\bibfnamefont {C.~D.}\ \bibnamefont {Malliakas}}, \bibinfo {author}
  {\bibfnamefont {J.~F.}\ \bibnamefont {Mitchell}}, \bibinfo {author}
  {\bibfnamefont {K.}~\bibnamefont {Mehlawat}}, \bibinfo {author}
  {\bibfnamefont {Y.}~\bibnamefont {Singh}}, \bibinfo {author} {\bibfnamefont
  {Y.}~\bibnamefont {Choi}}, \bibinfo {author} {\bibfnamefont {T.}~\bibnamefont
  {Gog}}, \bibinfo {author} {\bibfnamefont {A.}~\bibnamefont {Al-Zein}},
  \bibinfo {author} {\bibfnamefont {M.~M.}\ \bibnamefont {Sala}}, \bibinfo
  {author} {\bibfnamefont {M.}~\bibnamefont {Krisch}}, \bibinfo {author}
  {\bibfnamefont {J.}~\bibnamefont {Chaloupka}}, \bibinfo {author}
  {\bibfnamefont {G.}~\bibnamefont {Jackeli}}, \bibinfo {author} {\bibfnamefont
  {G.}~\bibnamefont {Khaliullin}}, \ and\ \bibinfo {author} {\bibfnamefont
  {B.~J.}\ \bibnamefont {Kim}},\ }\href {\doibase 10.1038/nphys3322} {\bibfield
   {journal} {\bibinfo  {journal} {Nat. Phys.}\ }\textbf {\bibinfo {volume}
  {11}},\ \bibinfo {pages} {462} (\bibinfo {year} {2015})}\BibitemShut
  {NoStop}%
\bibitem [{\citenamefont {Chaloupka}\ and\ \citenamefont
  {Khaliullin}(2016)}]{Chaloupka2016}%
  \BibitemOpen
  \bibfield  {author} {\bibinfo {author} {\bibfnamefont {J.}~\bibnamefont
  {Chaloupka}}\ and\ \bibinfo {author} {\bibfnamefont {G.}~\bibnamefont
  {Khaliullin}},\ }\href {\doibase 10.1103/PhysRevB.94.064435} {\bibfield
  {journal} {\bibinfo  {journal} {Phys. Rev. B}\ }\textbf {\bibinfo {volume}
  {94}},\ \bibinfo {pages} {064435} (\bibinfo {year} {2016})}\BibitemShut
  {NoStop}%
\bibitem [{\citenamefont {Yang}\ and\ \citenamefont {Kim}(2010)}]{Yang2010}%
  \BibitemOpen
  \bibfield  {author} {\bibinfo {author} {\bibfnamefont {B.-J.}\ \bibnamefont
  {Yang}}\ and\ \bibinfo {author} {\bibfnamefont {Y.~B.}\ \bibnamefont {Kim}},\
  }\href {\doibase 10.1103/PhysRevB.82.085111} {\bibfield  {journal} {\bibinfo
  {journal} {Phys. Rev. B}\ }\textbf {\bibinfo {volume} {82}},\ \bibinfo
  {pages} {085111} (\bibinfo {year} {2010})}\BibitemShut {NoStop}%
\bibitem [{\citenamefont {Wan}\ \emph {et~al.}(2011)\citenamefont {Wan},
  \citenamefont {Turner}, \citenamefont {Vishwanath},\ and\ \citenamefont
  {Savrasov}}]{Wan2011}%
  \BibitemOpen
  \bibfield  {author} {\bibinfo {author} {\bibfnamefont {X.}~\bibnamefont
  {Wan}}, \bibinfo {author} {\bibfnamefont {A.~M.}\ \bibnamefont {Turner}},
  \bibinfo {author} {\bibfnamefont {A.}~\bibnamefont {Vishwanath}}, \ and\
  \bibinfo {author} {\bibfnamefont {S.~Y.}\ \bibnamefont {Savrasov}},\ }\href
  {\doibase 10.1103/PhysRevB.83.205101} {\bibfield  {journal} {\bibinfo
  {journal} {Phys. Rev. B - Condens. Matter Mater. Phys.}\ }\textbf {\bibinfo
  {volume} {83}},\ \bibinfo {pages} {205101} (\bibinfo {year} {2011})},\
  \Eprint {http://arxiv.org/abs/1007.0016} {arXiv:1007.0016} \BibitemShut
  {NoStop}%
\bibitem [{\citenamefont {Witczak-Krempa}\ and\ \citenamefont
  {Kim}(2012)}]{Witczak-Krempa2012}%
  \BibitemOpen
  \bibfield  {author} {\bibinfo {author} {\bibfnamefont {W.}~\bibnamefont
  {Witczak-Krempa}}\ and\ \bibinfo {author} {\bibfnamefont {Y.~B.}\
  \bibnamefont {Kim}},\ }\href {\doibase 10.1103/PhysRevB.85.045124} {\bibfield
   {journal} {\bibinfo  {journal} {Phys. Rev. B}\ }\textbf {\bibinfo {volume}
  {85}},\ \bibinfo {pages} {045124} (\bibinfo {year} {2012})}\BibitemShut
  {NoStop}%
\bibitem [{\citenamefont {Donnerer}\ \emph {et~al.}(2016)\citenamefont
  {Donnerer}, \citenamefont {Rahn}, \citenamefont {Sala}, \citenamefont {Vale},
  \citenamefont {Pincini}, \citenamefont {Strempfer}, \citenamefont {Krisch},
  \citenamefont {Prabhakaran}, \citenamefont {Boothroyd},\ and\ \citenamefont
  {McMorrow}}]{Donnerer2016a}%
  \BibitemOpen
  \bibfield  {author} {\bibinfo {author} {\bibfnamefont {C.}~\bibnamefont
  {Donnerer}}, \bibinfo {author} {\bibfnamefont {M.~C.}\ \bibnamefont {Rahn}},
  \bibinfo {author} {\bibfnamefont {M.~M.}\ \bibnamefont {Sala}}, \bibinfo
  {author} {\bibfnamefont {J.~G.}\ \bibnamefont {Vale}}, \bibinfo {author}
  {\bibfnamefont {D.}~\bibnamefont {Pincini}}, \bibinfo {author} {\bibfnamefont
  {J.}~\bibnamefont {Strempfer}}, \bibinfo {author} {\bibfnamefont
  {M.}~\bibnamefont {Krisch}}, \bibinfo {author} {\bibfnamefont
  {D.}~\bibnamefont {Prabhakaran}}, \bibinfo {author} {\bibfnamefont {A.~T.}\
  \bibnamefont {Boothroyd}}, \ and\ \bibinfo {author} {\bibfnamefont {D.~F.}\
  \bibnamefont {McMorrow}},\ }\href {\doibase 10.1103/PhysRevLett.117.037201}
  {\bibfield  {journal} {\bibinfo  {journal} {Phys. Rev. Lett.}\ }\textbf
  {\bibinfo {volume} {117}},\ \bibinfo {pages} {037201} (\bibinfo {year}
  {2016})}\BibitemShut {NoStop}%
\bibitem [{\citenamefont {Crawford}\ \emph {et~al.}(1994)\citenamefont
  {Crawford}, \citenamefont {Subramanian}, \citenamefont {Harlow},
  \citenamefont {Fernandez-Baca}, \citenamefont {Wang},\ and\ \citenamefont
  {Johnston}}]{Crawford1994}%
  \BibitemOpen
  \bibfield  {author} {\bibinfo {author} {\bibfnamefont {M.~K.}\ \bibnamefont
  {Crawford}}, \bibinfo {author} {\bibfnamefont {M.~A.}\ \bibnamefont
  {Subramanian}}, \bibinfo {author} {\bibfnamefont {R.~L.}\ \bibnamefont
  {Harlow}}, \bibinfo {author} {\bibfnamefont {J.~A.}\ \bibnamefont
  {Fernandez-Baca}}, \bibinfo {author} {\bibfnamefont {Z.~R.}\ \bibnamefont
  {Wang}}, \ and\ \bibinfo {author} {\bibfnamefont {D.~C.}\ \bibnamefont
  {Johnston}},\ }\href {\doibase 10.1103/PhysRevB.49.9198} {\bibfield
  {journal} {\bibinfo  {journal} {Phys. Rev. B}\ }\textbf {\bibinfo {volume}
  {49}},\ \bibinfo {pages} {9198} (\bibinfo {year} {1994})}\BibitemShut
  {NoStop}%
\bibitem [{\citenamefont {Kim}\ \emph {et~al.}(2012)\citenamefont {Kim},
  \citenamefont {Casa}, \citenamefont {Upton}, \citenamefont {Gog},
  \citenamefont {Kim}, \citenamefont {Mitchell}, \citenamefont {van
  Veenendaal}, \citenamefont {Daghofer}, \citenamefont {van~den Brink},
  \citenamefont {Khaliullin},\ and\ \citenamefont {Kim}}]{Kim2012}%
  \BibitemOpen
  \bibfield  {author} {\bibinfo {author} {\bibfnamefont {J.}~\bibnamefont
  {Kim}}, \bibinfo {author} {\bibfnamefont {D.}~\bibnamefont {Casa}}, \bibinfo
  {author} {\bibfnamefont {M.~H.}\ \bibnamefont {Upton}}, \bibinfo {author}
  {\bibfnamefont {T.}~\bibnamefont {Gog}}, \bibinfo {author} {\bibfnamefont
  {Y.-J.}\ \bibnamefont {Kim}}, \bibinfo {author} {\bibfnamefont {J.~F.}\
  \bibnamefont {Mitchell}}, \bibinfo {author} {\bibfnamefont {M.}~\bibnamefont
  {van Veenendaal}}, \bibinfo {author} {\bibfnamefont {M.}~\bibnamefont
  {Daghofer}}, \bibinfo {author} {\bibfnamefont {J.}~\bibnamefont {van~den
  Brink}}, \bibinfo {author} {\bibfnamefont {G.}~\bibnamefont {Khaliullin}}, \
  and\ \bibinfo {author} {\bibfnamefont {B.~J.}\ \bibnamefont {Kim}},\ }\href
  {\doibase 10.1103/PhysRevLett.108.177003} {\bibfield  {journal} {\bibinfo
  {journal} {Phys. Rev. Lett.}\ }\textbf {\bibinfo {volume} {108}},\ \bibinfo
  {pages} {177003} (\bibinfo {year} {2012})},\ \Eprint
  {http://arxiv.org/abs/1110.0759} {arXiv:1110.0759} \BibitemShut {NoStop}%
\bibitem [{\citenamefont {Wang}\ and\ \citenamefont
  {Senthil}(2011)}]{Wang2011}%
  \BibitemOpen
  \bibfield  {author} {\bibinfo {author} {\bibfnamefont {F.}~\bibnamefont
  {Wang}}\ and\ \bibinfo {author} {\bibfnamefont {T.}~\bibnamefont {Senthil}},\
  }\href {\doibase 10.1103/PhysRevLett.106.136402} {\bibfield  {journal}
  {\bibinfo  {journal} {Phys. Rev. Lett.}\ }\textbf {\bibinfo {volume} {106}},\
  \bibinfo {pages} {136402} (\bibinfo {year} {2011})},\ \Eprint
  {http://arxiv.org/abs/1011.3500} {arXiv:1011.3500} \BibitemShut {NoStop}%
\bibitem [{\citenamefont {Watanabe}\ \emph {et~al.}(2013)\citenamefont
  {Watanabe}, \citenamefont {Shirakawa},\ and\ \citenamefont
  {Yunoki}}]{Watanabe2013}%
  \BibitemOpen
  \bibfield  {author} {\bibinfo {author} {\bibfnamefont {H.}~\bibnamefont
  {Watanabe}}, \bibinfo {author} {\bibfnamefont {T.}~\bibnamefont {Shirakawa}},
  \ and\ \bibinfo {author} {\bibfnamefont {S.}~\bibnamefont {Yunoki}},\ }\href
  {\doibase 10.1103/PhysRevLett.110.027002} {\bibfield  {journal} {\bibinfo
  {journal} {Phys. Rev. Lett.}\ }\textbf {\bibinfo {volume} {110}},\ \bibinfo
  {pages} {027002} (\bibinfo {year} {2013})}\BibitemShut {NoStop}%
\bibitem [{\citenamefont {Yang}\ \emph {et~al.}(2014)\citenamefont {Yang},
  \citenamefont {Wang}, \citenamefont {Liu}, \citenamefont {Chen},
  \citenamefont {Dai},\ and\ \citenamefont {Wang}}]{Yang2014}%
  \BibitemOpen
  \bibfield  {author} {\bibinfo {author} {\bibfnamefont {Y.}~\bibnamefont
  {Yang}}, \bibinfo {author} {\bibfnamefont {W.-S.}\ \bibnamefont {Wang}},
  \bibinfo {author} {\bibfnamefont {J.-G.}\ \bibnamefont {Liu}}, \bibinfo
  {author} {\bibfnamefont {H.}~\bibnamefont {Chen}}, \bibinfo {author}
  {\bibfnamefont {J.-H.}\ \bibnamefont {Dai}}, \ and\ \bibinfo {author}
  {\bibfnamefont {Q.-H.}\ \bibnamefont {Wang}},\ }\href {\doibase
  10.1103/PhysRevB.89.094518} {\bibfield  {journal} {\bibinfo  {journal} {Phys.
  Rev. B}\ }\textbf {\bibinfo {volume} {89}},\ \bibinfo {pages} {094518}
  (\bibinfo {year} {2014})},\ \Eprint {http://arxiv.org/abs/1312.4025}
  {arXiv:1312.4025} \BibitemShut {NoStop}%
\bibitem [{\citenamefont {Gretarsson}\ \emph {et~al.}(2016)\citenamefont
  {Gretarsson}, \citenamefont {Sung}, \citenamefont {Porras}, \citenamefont
  {Bertinshaw}, \citenamefont {Dietl}, \citenamefont {Bruin}, \citenamefont
  {Bangura}, \citenamefont {Kim}, \citenamefont {Dinnebier}, \citenamefont
  {Kim}, \citenamefont {Al-Zein}, \citenamefont {{Moretti Sala}}, \citenamefont
  {Krisch}, \citenamefont {{Le Tacon}}, \citenamefont {Keimer},\ and\
  \citenamefont {Kim}}]{Gretarsson2016a}%
  \BibitemOpen
  \bibfield  {author} {\bibinfo {author} {\bibfnamefont {H.}~\bibnamefont
  {Gretarsson}}, \bibinfo {author} {\bibfnamefont {N.~H.}\ \bibnamefont
  {Sung}}, \bibinfo {author} {\bibfnamefont {J.}~\bibnamefont {Porras}},
  \bibinfo {author} {\bibfnamefont {J.}~\bibnamefont {Bertinshaw}}, \bibinfo
  {author} {\bibfnamefont {C.}~\bibnamefont {Dietl}}, \bibinfo {author}
  {\bibfnamefont {J.~A.~N.}\ \bibnamefont {Bruin}}, \bibinfo {author}
  {\bibfnamefont {A.~F.}\ \bibnamefont {Bangura}}, \bibinfo {author}
  {\bibfnamefont {Y.~K.}\ \bibnamefont {Kim}}, \bibinfo {author} {\bibfnamefont
  {R.}~\bibnamefont {Dinnebier}}, \bibinfo {author} {\bibfnamefont
  {J.}~\bibnamefont {Kim}}, \bibinfo {author} {\bibfnamefont {A.}~\bibnamefont
  {Al-Zein}}, \bibinfo {author} {\bibfnamefont {M.}~\bibnamefont {{Moretti
  Sala}}}, \bibinfo {author} {\bibfnamefont {M.}~\bibnamefont {Krisch}},
  \bibinfo {author} {\bibfnamefont {M.}~\bibnamefont {{Le Tacon}}}, \bibinfo
  {author} {\bibfnamefont {B.}~\bibnamefont {Keimer}}, \ and\ \bibinfo {author}
  {\bibfnamefont {B.~J.}\ \bibnamefont {Kim}},\ }\href {\doibase
  10.1103/PhysRevLett.117.107001} {\bibfield  {journal} {\bibinfo  {journal}
  {Phys. Rev. Lett.}\ }\textbf {\bibinfo {volume} {117}},\ \bibinfo {pages}
  {107001} (\bibinfo {year} {2016})}\BibitemShut {NoStop}%
\bibitem [{\citenamefont {Tacon}\ \emph {et~al.}(2011)\citenamefont {Tacon},
  \citenamefont {Ghiringhelli}, \citenamefont {Chaloupka}, \citenamefont
  {Sala}, \citenamefont {Hinkov}, \citenamefont {Haverkort}, \citenamefont
  {Minola}, \citenamefont {Bakr}, \citenamefont {Zhou}, \citenamefont
  {Blanco-Canosa}, \citenamefont {Monney}, \citenamefont {Song}, \citenamefont
  {Sun}, \citenamefont {Lin}, \citenamefont {{De Luca}}, \citenamefont
  {Salluzzo}, \citenamefont {Khaliullin}, \citenamefont {Schmitt},
  \citenamefont {Braicovich},\ and\ \citenamefont {Keimer}}]{LeTacon2011}%
  \BibitemOpen
  \bibfield  {author} {\bibinfo {author} {\bibfnamefont {M.~L.}\ \bibnamefont
  {Tacon}}, \bibinfo {author} {\bibfnamefont {G.}~\bibnamefont {Ghiringhelli}},
  \bibinfo {author} {\bibfnamefont {J.}~\bibnamefont {Chaloupka}}, \bibinfo
  {author} {\bibfnamefont {M.~M.}\ \bibnamefont {Sala}}, \bibinfo {author}
  {\bibfnamefont {V.}~\bibnamefont {Hinkov}}, \bibinfo {author} {\bibfnamefont
  {M.~W.}\ \bibnamefont {Haverkort}}, \bibinfo {author} {\bibfnamefont
  {M.}~\bibnamefont {Minola}}, \bibinfo {author} {\bibfnamefont
  {M.}~\bibnamefont {Bakr}}, \bibinfo {author} {\bibfnamefont {K.~J.}\
  \bibnamefont {Zhou}}, \bibinfo {author} {\bibfnamefont {S.}~\bibnamefont
  {Blanco-Canosa}}, \bibinfo {author} {\bibfnamefont {C.}~\bibnamefont
  {Monney}}, \bibinfo {author} {\bibfnamefont {Y.~T.}\ \bibnamefont {Song}},
  \bibinfo {author} {\bibfnamefont {G.~L.}\ \bibnamefont {Sun}}, \bibinfo
  {author} {\bibfnamefont {C.~T.}\ \bibnamefont {Lin}}, \bibinfo {author}
  {\bibfnamefont {G.~M.}\ \bibnamefont {{De Luca}}}, \bibinfo {author}
  {\bibfnamefont {M.}~\bibnamefont {Salluzzo}}, \bibinfo {author}
  {\bibfnamefont {G.}~\bibnamefont {Khaliullin}}, \bibinfo {author}
  {\bibfnamefont {T.}~\bibnamefont {Schmitt}}, \bibinfo {author} {\bibfnamefont
  {L.}~\bibnamefont {Braicovich}}, \ and\ \bibinfo {author} {\bibfnamefont
  {B.}~\bibnamefont {Keimer}},\ }\href {\doibase 10.1038/nphys2041} {\bibfield
  {journal} {\bibinfo  {journal} {Nat. Phys.}\ }\textbf {\bibinfo {volume}
  {7}},\ \bibinfo {pages} {11} (\bibinfo {year} {2011})},\ \Eprint
  {http://arxiv.org/abs/1106.2641} {arXiv:1106.2641} \BibitemShut {NoStop}%
\bibitem [{\citenamefont {{Le Tacon}}\ \emph {et~al.}(2013)\citenamefont {{Le
  Tacon}}, \citenamefont {Minola}, \citenamefont {Peets}, \citenamefont
  {{Moretti Sala}}, \citenamefont {Blanco-Canosa}, \citenamefont {Hinkov},
  \citenamefont {Liang}, \citenamefont {Bonn}, \citenamefont {Hardy},
  \citenamefont {Lin}, \citenamefont {Schmitt}, \citenamefont {Braicovich},
  \citenamefont {Ghiringhelli},\ and\ \citenamefont {Keimer}}]{LeTacon2013}%
  \BibitemOpen
  \bibfield  {author} {\bibinfo {author} {\bibfnamefont {M.}~\bibnamefont {{Le
  Tacon}}}, \bibinfo {author} {\bibfnamefont {M.}~\bibnamefont {Minola}},
  \bibinfo {author} {\bibfnamefont {D.~C.}\ \bibnamefont {Peets}}, \bibinfo
  {author} {\bibfnamefont {M.}~\bibnamefont {{Moretti Sala}}}, \bibinfo
  {author} {\bibfnamefont {S.}~\bibnamefont {Blanco-Canosa}}, \bibinfo {author}
  {\bibfnamefont {V.}~\bibnamefont {Hinkov}}, \bibinfo {author} {\bibfnamefont
  {R.}~\bibnamefont {Liang}}, \bibinfo {author} {\bibfnamefont {D.~A.}\
  \bibnamefont {Bonn}}, \bibinfo {author} {\bibfnamefont {W.~N.}\ \bibnamefont
  {Hardy}}, \bibinfo {author} {\bibfnamefont {C.~T.}\ \bibnamefont {Lin}},
  \bibinfo {author} {\bibfnamefont {T.}~\bibnamefont {Schmitt}}, \bibinfo
  {author} {\bibfnamefont {L.}~\bibnamefont {Braicovich}}, \bibinfo {author}
  {\bibfnamefont {G.}~\bibnamefont {Ghiringhelli}}, \ and\ \bibinfo {author}
  {\bibfnamefont {B.}~\bibnamefont {Keimer}},\ }\href {\doibase
  10.1103/PhysRevB.88.020501} {\bibfield  {journal} {\bibinfo  {journal} {Phys.
  Rev. B}\ }\textbf {\bibinfo {volume} {88}},\ \bibinfo {pages} {020501}
  (\bibinfo {year} {2013})}\BibitemShut {NoStop}%
\bibitem [{\citenamefont {Dean}\ \emph {et~al.}(2013)\citenamefont {Dean},
  \citenamefont {Dellea}, \citenamefont {Springell}, \citenamefont
  {Yakhou-Harris}, \citenamefont {Kummer}, \citenamefont {Brookes},
  \citenamefont {Liu}, \citenamefont {Sun}, \citenamefont {Strle},
  \citenamefont {Schmitt}, \citenamefont {Braicovich}, \citenamefont
  {Ghiringhelli}, \citenamefont {Bo{\v{z}}ovi{\'{c}}},\ and\ \citenamefont
  {Hill}}]{Dean2013}%
  \BibitemOpen
  \bibfield  {author} {\bibinfo {author} {\bibfnamefont {M.~P.~M.}\
  \bibnamefont {Dean}}, \bibinfo {author} {\bibfnamefont {G.}~\bibnamefont
  {Dellea}}, \bibinfo {author} {\bibfnamefont {R.~S.}\ \bibnamefont
  {Springell}}, \bibinfo {author} {\bibfnamefont {F.}~\bibnamefont
  {Yakhou-Harris}}, \bibinfo {author} {\bibfnamefont {K.}~\bibnamefont
  {Kummer}}, \bibinfo {author} {\bibfnamefont {N.~B.}\ \bibnamefont {Brookes}},
  \bibinfo {author} {\bibfnamefont {X.}~\bibnamefont {Liu}}, \bibinfo {author}
  {\bibfnamefont {Y.-J.}\ \bibnamefont {Sun}}, \bibinfo {author} {\bibfnamefont
  {J.}~\bibnamefont {Strle}}, \bibinfo {author} {\bibfnamefont
  {T.}~\bibnamefont {Schmitt}}, \bibinfo {author} {\bibfnamefont
  {L.}~\bibnamefont {Braicovich}}, \bibinfo {author} {\bibfnamefont
  {G.}~\bibnamefont {Ghiringhelli}}, \bibinfo {author} {\bibfnamefont
  {I.}~\bibnamefont {Bo{\v{z}}ovi{\'{c}}}}, \ and\ \bibinfo {author}
  {\bibfnamefont {J.~P.}\ \bibnamefont {Hill}},\ }\href {\doibase
  10.1038/nmat3723} {\bibfield  {journal} {\bibinfo  {journal} {Nat. Mater.}\
  }\textbf {\bibinfo {volume} {12}},\ \bibinfo {pages} {1019} (\bibinfo {year}
  {2013})},\ \Eprint {http://arxiv.org/abs/1303.5359} {arXiv:1303.5359}
  \BibitemShut {NoStop}%
\bibitem [{\citenamefont {Kim}\ \emph {et~al.}(2014{\natexlab{a}})\citenamefont
  {Kim}, \citenamefont {Krupin}, \citenamefont {Denlinger}, \citenamefont
  {Bostwick}, \citenamefont {Rotenberg}, \citenamefont {Zhao}, \citenamefont
  {Mitchell}, \citenamefont {Allen},\ and\ \citenamefont {Kim}}]{Kim2014a}%
  \BibitemOpen
  \bibfield  {author} {\bibinfo {author} {\bibfnamefont {Y.~K.}\ \bibnamefont
  {Kim}}, \bibinfo {author} {\bibfnamefont {O.}~\bibnamefont {Krupin}},
  \bibinfo {author} {\bibfnamefont {J.~D.}\ \bibnamefont {Denlinger}}, \bibinfo
  {author} {\bibfnamefont {A.}~\bibnamefont {Bostwick}}, \bibinfo {author}
  {\bibfnamefont {E.}~\bibnamefont {Rotenberg}}, \bibinfo {author}
  {\bibfnamefont {Q.}~\bibnamefont {Zhao}}, \bibinfo {author} {\bibfnamefont
  {J.~F.}\ \bibnamefont {Mitchell}}, \bibinfo {author} {\bibfnamefont {J.~W.}\
  \bibnamefont {Allen}}, \ and\ \bibinfo {author} {\bibfnamefont {B.~J.}\
  \bibnamefont {Kim}},\ }\href {\doibase 10.1126/science.1251151} {\bibfield
  {journal} {\bibinfo  {journal} {Science (80-. ).}\ }\textbf {\bibinfo
  {volume} {345}},\ \bibinfo {pages} {187} (\bibinfo {year}
  {2014}{\natexlab{a}})}\BibitemShut {NoStop}%
\bibitem [{\citenamefont {de~la Torre}\ \emph {et~al.}(2015)\citenamefont
  {de~la Torre}, \citenamefont {{McKeown Walker}}, \citenamefont {Bruno},
  \citenamefont {Ricc{\'{o}}}, \citenamefont {Wang}, \citenamefont {{Gutierrez
  Lezama}}, \citenamefont {Scheerer}, \citenamefont {Giriat}, \citenamefont
  {Jaccard}, \citenamefont {Berthod}, \citenamefont {Kim}, \citenamefont
  {Hoesch}, \citenamefont {Hunter}, \citenamefont {Perry}, \citenamefont
  {Tamai},\ and\ \citenamefont {Baumberger}}]{DeLaTorre2015}%
  \BibitemOpen
  \bibfield  {author} {\bibinfo {author} {\bibfnamefont {A.}~\bibnamefont
  {de~la Torre}}, \bibinfo {author} {\bibfnamefont {S.}~\bibnamefont {{McKeown
  Walker}}}, \bibinfo {author} {\bibfnamefont {F.~Y.}\ \bibnamefont {Bruno}},
  \bibinfo {author} {\bibfnamefont {S.}~\bibnamefont {Ricc{\'{o}}}}, \bibinfo
  {author} {\bibfnamefont {Z.}~\bibnamefont {Wang}}, \bibinfo {author}
  {\bibfnamefont {I.}~\bibnamefont {{Gutierrez Lezama}}}, \bibinfo {author}
  {\bibfnamefont {G.}~\bibnamefont {Scheerer}}, \bibinfo {author}
  {\bibfnamefont {G.}~\bibnamefont {Giriat}}, \bibinfo {author} {\bibfnamefont
  {D.}~\bibnamefont {Jaccard}}, \bibinfo {author} {\bibfnamefont
  {C.}~\bibnamefont {Berthod}}, \bibinfo {author} {\bibfnamefont {T.~K.}\
  \bibnamefont {Kim}}, \bibinfo {author} {\bibfnamefont {M.}~\bibnamefont
  {Hoesch}}, \bibinfo {author} {\bibfnamefont {E.~C.}\ \bibnamefont {Hunter}},
  \bibinfo {author} {\bibfnamefont {R.~S.}\ \bibnamefont {Perry}}, \bibinfo
  {author} {\bibfnamefont {A.}~\bibnamefont {Tamai}}, \ and\ \bibinfo {author}
  {\bibfnamefont {F.}~\bibnamefont {Baumberger}},\ }\href {\doibase
  10.1103/PhysRevLett.115.176402} {\bibfield  {journal} {\bibinfo  {journal}
  {Phys. Rev. Lett.}\ }\textbf {\bibinfo {volume} {115}},\ \bibinfo {pages}
  {176402} (\bibinfo {year} {2015})}\BibitemShut {NoStop}%
\bibitem [{\citenamefont {Kim}\ \emph {et~al.}(2015)\citenamefont {Kim},
  \citenamefont {Sung}, \citenamefont {Denlinger},\ and\ \citenamefont
  {Kim}}]{Kim2016}%
  \BibitemOpen
  \bibfield  {author} {\bibinfo {author} {\bibfnamefont {Y.~K.}\ \bibnamefont
  {Kim}}, \bibinfo {author} {\bibfnamefont {N.~H.}\ \bibnamefont {Sung}},
  \bibinfo {author} {\bibfnamefont {J.~D.}\ \bibnamefont {Denlinger}}, \ and\
  \bibinfo {author} {\bibfnamefont {B.~J.}\ \bibnamefont {Kim}},\ }\href
  {\doibase 10.1038/nphys3503} {\bibfield  {journal} {\bibinfo  {journal} {Nat.
  Phys.}\ }\textbf {\bibinfo {volume} {12}},\ \bibinfo {pages} {37} (\bibinfo
  {year} {2015})}\BibitemShut {NoStop}%
\bibitem [{\citenamefont {Birol}\ and\ \citenamefont
  {Haule}(2015)}]{Birol2015}%
  \BibitemOpen
  \bibfield  {author} {\bibinfo {author} {\bibfnamefont {T.}~\bibnamefont
  {Birol}}\ and\ \bibinfo {author} {\bibfnamefont {K.}~\bibnamefont {Haule}},\
  }\href {\doibase 10.1103/PhysRevLett.114.096403} {\bibfield  {journal}
  {\bibinfo  {journal} {Phys. Rev. Lett.}\ }\textbf {\bibinfo {volume} {114}},\
  \bibinfo {pages} {096403} (\bibinfo {year} {2015})}\BibitemShut {NoStop}%
\bibitem [{\citenamefont {Pedersen}\ \emph {et~al.}(2016)\citenamefont
  {Pedersen}, \citenamefont {Bendix}, \citenamefont {Tressaud}, \citenamefont
  {Durand}, \citenamefont {Weihe}, \citenamefont {Salman}, \citenamefont
  {Morsing}, \citenamefont {Woodruff}, \citenamefont {Lan}, \citenamefont
  {Wernsdorfer}, \citenamefont {Mathoni{\`{e}}re}, \citenamefont {Piligkos},
  \citenamefont {Klokishner}, \citenamefont {Ostrovsky}, \citenamefont
  {Ollefs}, \citenamefont {Wilhelm}, \citenamefont {Rogalev},\ and\
  \citenamefont {Cl{\'{e}}rac}}]{Pedersen2016}%
  \BibitemOpen
  \bibfield  {author} {\bibinfo {author} {\bibfnamefont {K.~S.}\ \bibnamefont
  {Pedersen}}, \bibinfo {author} {\bibfnamefont {J.}~\bibnamefont {Bendix}},
  \bibinfo {author} {\bibfnamefont {A.}~\bibnamefont {Tressaud}}, \bibinfo
  {author} {\bibfnamefont {E.}~\bibnamefont {Durand}}, \bibinfo {author}
  {\bibfnamefont {H.}~\bibnamefont {Weihe}}, \bibinfo {author} {\bibfnamefont
  {Z.}~\bibnamefont {Salman}}, \bibinfo {author} {\bibfnamefont {T.~J.}\
  \bibnamefont {Morsing}}, \bibinfo {author} {\bibfnamefont {D.~N.}\
  \bibnamefont {Woodruff}}, \bibinfo {author} {\bibfnamefont {Y.}~\bibnamefont
  {Lan}}, \bibinfo {author} {\bibfnamefont {W.}~\bibnamefont {Wernsdorfer}},
  \bibinfo {author} {\bibfnamefont {C.}~\bibnamefont {Mathoni{\`{e}}re}},
  \bibinfo {author} {\bibfnamefont {S.}~\bibnamefont {Piligkos}}, \bibinfo
  {author} {\bibfnamefont {S.~I.}\ \bibnamefont {Klokishner}}, \bibinfo
  {author} {\bibfnamefont {S.}~\bibnamefont {Ostrovsky}}, \bibinfo {author}
  {\bibfnamefont {K.}~\bibnamefont {Ollefs}}, \bibinfo {author} {\bibfnamefont
  {F.}~\bibnamefont {Wilhelm}}, \bibinfo {author} {\bibfnamefont
  {A.}~\bibnamefont {Rogalev}}, \ and\ \bibinfo {author} {\bibfnamefont
  {R.}~\bibnamefont {Cl{\'{e}}rac}},\ }\href {\doibase 10.1038/ncomms12195}
  {\bibfield  {journal} {\bibinfo  {journal} {Nat. Commun.}\ }\textbf {\bibinfo
  {volume} {7}},\ \bibinfo {pages} {12195} (\bibinfo {year}
  {2016})}\BibitemShut {NoStop}%
\bibitem [{\citenamefont {Ponchut}\ \emph {et~al.}(2011)\citenamefont
  {Ponchut}, \citenamefont {Rigal}, \citenamefont {Cl{\'{e}}ment},
  \citenamefont {Papillon}, \citenamefont {Homs},\ and\ \citenamefont
  {Petitdemange}}]{Ponchut2011}%
  \BibitemOpen
  \bibfield  {author} {\bibinfo {author} {\bibfnamefont {C.}~\bibnamefont
  {Ponchut}}, \bibinfo {author} {\bibfnamefont {J.~M.}\ \bibnamefont {Rigal}},
  \bibinfo {author} {\bibfnamefont {J.}~\bibnamefont {Cl{\'{e}}ment}}, \bibinfo
  {author} {\bibfnamefont {E.}~\bibnamefont {Papillon}}, \bibinfo {author}
  {\bibfnamefont {A.}~\bibnamefont {Homs}}, \ and\ \bibinfo {author}
  {\bibfnamefont {S.}~\bibnamefont {Petitdemange}},\ }\href {\doibase
  10.1088/1748-0221/6/01/C01069} {\bibfield  {journal} {\bibinfo  {journal} {J.
  Instrum.}\ }\textbf {\bibinfo {volume} {6}},\ \bibinfo {pages} {C01069}
  (\bibinfo {year} {2011})}\BibitemShut {NoStop}%
\bibitem [{\citenamefont {{Moretti Sala}}\ \emph {et~al.}(2013)\citenamefont
  {{Moretti Sala}}, \citenamefont {Henriquet}, \citenamefont {Simonelli},
  \citenamefont {Verbeni},\ and\ \citenamefont {Monaco}}]{MorettiSala2013}%
  \BibitemOpen
  \bibfield  {author} {\bibinfo {author} {\bibfnamefont {M.}~\bibnamefont
  {{Moretti Sala}}}, \bibinfo {author} {\bibfnamefont {C.}~\bibnamefont
  {Henriquet}}, \bibinfo {author} {\bibfnamefont {L.}~\bibnamefont
  {Simonelli}}, \bibinfo {author} {\bibfnamefont {R.}~\bibnamefont {Verbeni}},
  \ and\ \bibinfo {author} {\bibfnamefont {G.}~\bibnamefont {Monaco}},\ }\href
  {\doibase 10.1016/j.elspec.2012.08.002} {\bibfield  {journal} {\bibinfo
  {journal} {J. Electron Spectros. Relat. Phenomena}\ }\textbf {\bibinfo
  {volume} {188}},\ \bibinfo {pages} {150} (\bibinfo {year}
  {2013})}\BibitemShut {NoStop}%
\bibitem [{\citenamefont {Neese}(2012)}]{Neese2012}%
  \BibitemOpen
  \bibfield  {author} {\bibinfo {author} {\bibfnamefont {F.}~\bibnamefont
  {Neese}},\ }\href {\doibase 10.1002/wcms.81} {\bibfield  {journal} {\bibinfo
  {journal} {Wiley Interdisciplinary Reviews: Computational Molecular Science}\
  }\textbf {\bibinfo {volume} {2}},\ \bibinfo {pages} {73} (\bibinfo {year}
  {2012})}\BibitemShut {NoStop}%
\bibitem [{\citenamefont {Angeli}\ \emph {et~al.}(2001)\citenamefont {Angeli},
  \citenamefont {Cimiraglia}, \citenamefont {Evangelisti}, \citenamefont
  {Leininger},\ and\ \citenamefont {Malrieu}}]{Angeli2001}%
  \BibitemOpen
  \bibfield  {author} {\bibinfo {author} {\bibfnamefont {C.}~\bibnamefont
  {Angeli}}, \bibinfo {author} {\bibfnamefont {R.}~\bibnamefont {Cimiraglia}},
  \bibinfo {author} {\bibfnamefont {S.}~\bibnamefont {Evangelisti}}, \bibinfo
  {author} {\bibfnamefont {T.}~\bibnamefont {Leininger}}, \ and\ \bibinfo
  {author} {\bibfnamefont {J.-P.}\ \bibnamefont {Malrieu}},\ }\href {\doibase
  10.1063/1.1361246} {\bibfield  {journal} {\bibinfo  {journal} {The Journal of
  Chemical Physics}\ }\textbf {\bibinfo {volume} {114}},\ \bibinfo {pages}
  {10252} (\bibinfo {year} {2001})}\BibitemShut {NoStop}%
\bibitem [{\citenamefont {Atanasov}\ \emph {et~al.}(2011)\citenamefont
  {Atanasov}, \citenamefont {Ganyushin}, \citenamefont {Pantazis},
  \citenamefont {Sivalingam},\ and\ \citenamefont {Neese}}]{Atanasov2011}%
  \BibitemOpen
  \bibfield  {author} {\bibinfo {author} {\bibfnamefont {M.}~\bibnamefont
  {Atanasov}}, \bibinfo {author} {\bibfnamefont {D.}~\bibnamefont {Ganyushin}},
  \bibinfo {author} {\bibfnamefont {D.~A.}\ \bibnamefont {Pantazis}}, \bibinfo
  {author} {\bibfnamefont {K.}~\bibnamefont {Sivalingam}}, \ and\ \bibinfo
  {author} {\bibfnamefont {F.}~\bibnamefont {Neese}},\ }\href {\doibase
  10.1021/ic200196k} {\bibfield  {journal} {\bibinfo  {journal} {Inorganic
  Chemistry}\ }\textbf {\bibinfo {volume} {50}},\ \bibinfo {pages} {7460}
  (\bibinfo {year} {2011})}\BibitemShut {NoStop}%
\bibitem [{\citenamefont {Ganyushin}\ and\ \citenamefont
  {Neese}(2006)}]{Ganyushin2006}%
  \BibitemOpen
  \bibfield  {author} {\bibinfo {author} {\bibfnamefont {D.}~\bibnamefont
  {Ganyushin}}\ and\ \bibinfo {author} {\bibfnamefont {F.}~\bibnamefont
  {Neese}},\ }\href {\doibase 10.1063/1.2213976} {\bibfield  {journal}
  {\bibinfo  {journal} {The Journal of Chemical Physics}\ }\textbf {\bibinfo
  {volume} {125}},\ \bibinfo {pages} {024103} (\bibinfo {year}
  {2006})}\BibitemShut {NoStop}%
\bibitem [{\citenamefont {Neese}(2005)}]{Neese2005}%
  \BibitemOpen
  \bibfield  {author} {\bibinfo {author} {\bibfnamefont {F.}~\bibnamefont
  {Neese}},\ }\href {\doibase 10.1063/1.1829047} {\bibfield  {journal}
  {\bibinfo  {journal} {The Journal of Chemical Physics}\ }\textbf {\bibinfo
  {volume} {122}},\ \bibinfo {pages} {034107} (\bibinfo {year}
  {2005})}\BibitemShut {NoStop}%
\bibitem [{\citenamefont {van Wüllen}(1998)}]{vanWullen1998}%
  \BibitemOpen
  \bibfield  {author} {\bibinfo {author} {\bibfnamefont {C.}~\bibnamefont {van
  Wüllen}},\ }\href {\doibase 10.1063/1.476576} {\bibfield  {journal}
  {\bibinfo  {journal} {The Journal of Chemical Physics}\ }\textbf {\bibinfo
  {volume} {109}},\ \bibinfo {pages} {392} (\bibinfo {year}
  {1998})}\BibitemShut {NoStop}%
\bibitem [{\citenamefont {Weigend}\ and\ \citenamefont
  {Ahlrichs}(2005)}]{Weigend2005}%
  \BibitemOpen
  \bibfield  {author} {\bibinfo {author} {\bibfnamefont {F.}~\bibnamefont
  {Weigend}}\ and\ \bibinfo {author} {\bibfnamefont {R.}~\bibnamefont
  {Ahlrichs}},\ }\href {\doibase 10.1039/b508541a} {\bibfield  {journal}
  {\bibinfo  {journal} {Physical Chemistry Chemical Physics}\ }\textbf
  {\bibinfo {volume} {7}},\ \bibinfo {pages} {3297} (\bibinfo {year}
  {2005})}\BibitemShut {NoStop}%
\bibitem [{\citenamefont {Pantazis}\ \emph {et~al.}(2008)\citenamefont
  {Pantazis}, \citenamefont {Chen}, \citenamefont {Landis},\ and\ \citenamefont
  {Neese}}]{Pantazis2008}%
  \BibitemOpen
  \bibfield  {author} {\bibinfo {author} {\bibfnamefont {D.~A.}\ \bibnamefont
  {Pantazis}}, \bibinfo {author} {\bibfnamefont {X.-Y.}\ \bibnamefont {Chen}},
  \bibinfo {author} {\bibfnamefont {C.~R.}\ \bibnamefont {Landis}}, \ and\
  \bibinfo {author} {\bibfnamefont {F.}~\bibnamefont {Neese}},\ }\href
  {\doibase 10.1021/ct800047t} {\bibfield  {journal} {\bibinfo  {journal}
  {Journal of Chemical Theory and Computation}\ }\textbf {\bibinfo {volume}
  {4}},\ \bibinfo {pages} {908} (\bibinfo {year} {2008})}\BibitemShut {NoStop}%
\bibitem [{\citenamefont {Staemmler}(2005)}]{Staemmler2005}%
  \BibitemOpen
  \bibfield  {author} {\bibinfo {author} {\bibfnamefont {V.}~\bibnamefont
  {Staemmler}},\ }in\ \href@noop {} {\emph {\bibinfo {booktitle} {Theoretical
  Aspects of Transition Metal Catalysis}}},\ Vol.~\bibinfo {volume} {12},\
  \bibinfo {editor} {edited by\ \bibinfo {editor} {\bibfnamefont
  {G.}~\bibnamefont {Frenking}}}\ (\bibinfo  {publisher} {Springer Berlin
  Heidelberg},\ \bibinfo {year} {2005})\ pp.\ \bibinfo {pages}
  {219--256}\BibitemShut {NoStop}%
\bibitem [{\citenamefont {Smolentsev}\ \emph
  {et~al.}(2007{\natexlab{a}})\citenamefont {Smolentsev}, \citenamefont
  {Gubanov}, \citenamefont {Naumov},\ and\ \citenamefont
  {Danilenko}}]{Smolentsev2007}%
  \BibitemOpen
  \bibfield  {author} {\bibinfo {author} {\bibfnamefont {A.~I.}\ \bibnamefont
  {Smolentsev}}, \bibinfo {author} {\bibfnamefont {A.~I.}\ \bibnamefont
  {Gubanov}}, \bibinfo {author} {\bibfnamefont {D.~Y.}\ \bibnamefont {Naumov}},
  \ and\ \bibinfo {author} {\bibfnamefont {A.~M.}\ \bibnamefont {Danilenko}},\
  }\href {\doibase 10.1107/S1600536807059995} {\bibfield  {journal} {\bibinfo
  {journal} {Acta Crystallogr. Sect. E Struct. Reports Online}\ }\textbf
  {\bibinfo {volume} {63}},\ \bibinfo {pages} {i200} (\bibinfo {year}
  {2007}{\natexlab{a}})}\BibitemShut {NoStop}%
\bibitem [{\citenamefont {Coelho}(2007)}]{Coelho2007}%
  \BibitemOpen
  \bibfield  {author} {\bibinfo {author} {\bibfnamefont {A.}~\bibnamefont
  {Coelho}},\ }in\ \href@noop {} {\emph {\bibinfo {booktitle} {{TOPAS Academic:
  General Profile and Structure Analysis Software for Powder Diffractrion
  Data}}}}\ (\bibinfo {address} {Bruker AXS, Karlsruhe},\ \bibinfo {year}
  {2007})\BibitemShut {NoStop}%
\bibitem [{\citenamefont {Balzar}\ \emph {et~al.}(2004)\citenamefont {Balzar},
  \citenamefont {Audebrand}, \citenamefont {Daymond}, \citenamefont {Fitch},
  \citenamefont {Hewat}, \citenamefont {Langford}, \citenamefont {{Le Bail}},
  \citenamefont {Lou{\"{e}}r}, \citenamefont {Masson}, \citenamefont {McCowan},
  \citenamefont {Popa}, \citenamefont {Stephens},\ and\ \citenamefont
  {Toby}}]{Balzar2004}%
  \BibitemOpen
  \bibfield  {author} {\bibinfo {author} {\bibfnamefont {D.}~\bibnamefont
  {Balzar}}, \bibinfo {author} {\bibfnamefont {N.}~\bibnamefont {Audebrand}},
  \bibinfo {author} {\bibfnamefont {M.~R.}\ \bibnamefont {Daymond}}, \bibinfo
  {author} {\bibfnamefont {A.}~\bibnamefont {Fitch}}, \bibinfo {author}
  {\bibfnamefont {A.}~\bibnamefont {Hewat}}, \bibinfo {author} {\bibfnamefont
  {J.~I.}\ \bibnamefont {Langford}}, \bibinfo {author} {\bibfnamefont
  {A.}~\bibnamefont {{Le Bail}}}, \bibinfo {author} {\bibfnamefont
  {D.}~\bibnamefont {Lou{\"{e}}r}}, \bibinfo {author} {\bibfnamefont
  {O.}~\bibnamefont {Masson}}, \bibinfo {author} {\bibfnamefont {C.~N.}\
  \bibnamefont {McCowan}}, \bibinfo {author} {\bibfnamefont {N.~C.}\
  \bibnamefont {Popa}}, \bibinfo {author} {\bibfnamefont {P.~W.}\ \bibnamefont
  {Stephens}}, \ and\ \bibinfo {author} {\bibfnamefont {B.~H.}\ \bibnamefont
  {Toby}},\ }\href {\doibase 10.1107/S0021889804022551} {\bibfield  {journal}
  {\bibinfo  {journal} {J. Appl. Crystallogr.}\ }\textbf {\bibinfo {volume}
  {37}},\ \bibinfo {pages} {911} (\bibinfo {year} {2004})}\BibitemShut
  {NoStop}%
\bibitem [{\citenamefont {Avram}\ and\ \citenamefont {Brik}(2013)}]{Avram2013}%
  \BibitemOpen
  \bibinfo {editor} {\bibfnamefont {N.~M.}\ \bibnamefont {Avram}}\ and\
  \bibinfo {editor} {\bibfnamefont {M.~G.}\ \bibnamefont {Brik}},\ eds.,\ \href
  {\doibase 10.1007/978-3-642-30838-3} {\emph {\bibinfo {title} {{Optical
  Properties of 3d-Ions in Crystals: Spectroscopy and Crystal Field
  Analysis}}}}\ (\bibinfo  {publisher} {Springer Berlin Heidelberg},\ \bibinfo
  {address} {Berlin, Heidelberg},\ \bibinfo {year} {2013})\BibitemShut
  {NoStop}%
\bibitem [{\citenamefont {Smolentsev}\ \emph
  {et~al.}(2007{\natexlab{b}})\citenamefont {Smolentsev}, \citenamefont
  {Gubanov}, \citenamefont {Naumov},\ and\ \citenamefont
  {Danilenko}}]{Smolentsev2007a}%
  \BibitemOpen
  \bibfield  {author} {\bibinfo {author} {\bibfnamefont {A.~I.}\ \bibnamefont
  {Smolentsev}}, \bibinfo {author} {\bibfnamefont {A.~I.}\ \bibnamefont
  {Gubanov}}, \bibinfo {author} {\bibfnamefont {D.~Y.}\ \bibnamefont {Naumov}},
  \ and\ \bibinfo {author} {\bibfnamefont {A.~M.}\ \bibnamefont {Danilenko}},\
  }\href {\doibase 10.1107/S160053680706000X} {\bibfield  {journal} {\bibinfo
  {journal} {Acta Crystallogr. Sect. E Struct. Reports Online}\ }\textbf
  {\bibinfo {volume} {63}},\ \bibinfo {pages} {i201} (\bibinfo {year}
  {2007}{\natexlab{b}})}\BibitemShut {NoStop}%
\bibitem [{\citenamefont {Fitz}\ \emph {et~al.}(2002)\citenamefont {Fitz},
  \citenamefont {M{\"{u}}ller}, \citenamefont {Graudejus},\ and\ \citenamefont
  {Bartlett}}]{Fitz2002}%
  \BibitemOpen
  \bibfield  {author} {\bibinfo {author} {\bibfnamefont {H.}~\bibnamefont
  {Fitz}}, \bibinfo {author} {\bibfnamefont {B.~G.}\ \bibnamefont
  {M{\"{u}}ller}}, \bibinfo {author} {\bibfnamefont {O.}~\bibnamefont
  {Graudejus}}, \ and\ \bibinfo {author} {\bibfnamefont {N.}~\bibnamefont
  {Bartlett}},\ }\href {\doibase
  10.1002/1521-3749(200201)628:1<133::AID-ZAAC133>3.0.CO;2-S} {\bibfield
  {journal} {\bibinfo  {journal} {Zeitschrift f{\"{u}}r Anorg. und Allg.
  Chemie}\ }\textbf {\bibinfo {volume} {628}},\ \bibinfo {pages} {133}
  (\bibinfo {year} {2002})}\BibitemShut {NoStop}%
\bibitem [{\citenamefont {Smolentsev}\ \emph
  {et~al.}(2007{\natexlab{c}})\citenamefont {Smolentsev}, \citenamefont
  {Gubanov},\ and\ \citenamefont {Danilenko}}]{Smolentsev2007b}%
  \BibitemOpen
  \bibfield  {author} {\bibinfo {author} {\bibfnamefont {A.~I.}\ \bibnamefont
  {Smolentsev}}, \bibinfo {author} {\bibfnamefont {A.~I.}\ \bibnamefont
  {Gubanov}}, \ and\ \bibinfo {author} {\bibfnamefont {A.~M.}\ \bibnamefont
  {Danilenko}},\ }\href {\doibase 10.1107/S0108270107044046} {\bibfield
  {journal} {\bibinfo  {journal} {Acta Crystallogr. Sect. C Cryst. Struct.
  Commun.}\ }\textbf {\bibinfo {volume} {63}},\ \bibinfo {pages} {i99}
  (\bibinfo {year} {2007}{\natexlab{c}})}\BibitemShut {NoStop}%
\bibitem [{\citenamefont {Ishii}\ \emph {et~al.}(2011)\citenamefont {Ishii},
  \citenamefont {Jarrige}, \citenamefont {Yoshida}, \citenamefont {Ikeuchi},
  \citenamefont {Mizuki}, \citenamefont {Ohashi}, \citenamefont {Takayama},
  \citenamefont {Matsuno},\ and\ \citenamefont {Takagi}}]{Ishii2011}%
  \BibitemOpen
  \bibfield  {author} {\bibinfo {author} {\bibfnamefont {K.}~\bibnamefont
  {Ishii}}, \bibinfo {author} {\bibfnamefont {I.}~\bibnamefont {Jarrige}},
  \bibinfo {author} {\bibfnamefont {M.}~\bibnamefont {Yoshida}}, \bibinfo
  {author} {\bibfnamefont {K.}~\bibnamefont {Ikeuchi}}, \bibinfo {author}
  {\bibfnamefont {J.}~\bibnamefont {Mizuki}}, \bibinfo {author} {\bibfnamefont
  {K.}~\bibnamefont {Ohashi}}, \bibinfo {author} {\bibfnamefont
  {T.}~\bibnamefont {Takayama}}, \bibinfo {author} {\bibfnamefont
  {J.}~\bibnamefont {Matsuno}}, \ and\ \bibinfo {author} {\bibfnamefont
  {H.}~\bibnamefont {Takagi}},\ }\href {\doibase 10.1103/PhysRevB.83.115121}
  {\bibfield  {journal} {\bibinfo  {journal} {Phys. Rev. B}\ }\textbf {\bibinfo
  {volume} {83}},\ \bibinfo {pages} {115121} (\bibinfo {year}
  {2011})}\BibitemShut {NoStop}%
\bibitem [{\citenamefont {{Moretti Sala}}\ \emph
  {et~al.}(2014{\natexlab{a}})\citenamefont {{Moretti Sala}}, \citenamefont
  {Ohgushi}, \citenamefont {Al-Zein}, \citenamefont {Hirata}, \citenamefont
  {Monaco},\ and\ \citenamefont {Krisch}}]{MorettiSala2014b}%
  \BibitemOpen
  \bibfield  {author} {\bibinfo {author} {\bibfnamefont {M.}~\bibnamefont
  {{Moretti Sala}}}, \bibinfo {author} {\bibfnamefont {K.}~\bibnamefont
  {Ohgushi}}, \bibinfo {author} {\bibfnamefont {A.}~\bibnamefont {Al-Zein}},
  \bibinfo {author} {\bibfnamefont {Y.}~\bibnamefont {Hirata}}, \bibinfo
  {author} {\bibfnamefont {G.}~\bibnamefont {Monaco}}, \ and\ \bibinfo {author}
  {\bibfnamefont {M.}~\bibnamefont {Krisch}},\ }\href {\doibase
  10.1103/PhysRevLett.112.176402} {\bibfield  {journal} {\bibinfo  {journal}
  {Phys. Rev. Lett.}\ }\textbf {\bibinfo {volume} {112}},\ \bibinfo {pages}
  {176402} (\bibinfo {year} {2014}{\natexlab{a}})}\BibitemShut {NoStop}%
\bibitem [{\citenamefont {Lefran{\c{c}}ois}\ \emph {et~al.}(2016)\citenamefont
  {Lefran{\c{c}}ois}, \citenamefont {Pradipto}, \citenamefont {{Moretti Sala}},
  \citenamefont {Chapon}, \citenamefont {Simonet}, \citenamefont {Picozzi},
  \citenamefont {Lejay}, \citenamefont {Petit},\ and\ \citenamefont
  {Ballou}}]{Lefrancois2016}%
  \BibitemOpen
  \bibfield  {author} {\bibinfo {author} {\bibfnamefont {E.}~\bibnamefont
  {Lefran{\c{c}}ois}}, \bibinfo {author} {\bibfnamefont {A.-M.}\ \bibnamefont
  {Pradipto}}, \bibinfo {author} {\bibfnamefont {M.}~\bibnamefont {{Moretti
  Sala}}}, \bibinfo {author} {\bibfnamefont {L.~C.}\ \bibnamefont {Chapon}},
  \bibinfo {author} {\bibfnamefont {V.}~\bibnamefont {Simonet}}, \bibinfo
  {author} {\bibfnamefont {S.}~\bibnamefont {Picozzi}}, \bibinfo {author}
  {\bibfnamefont {P.}~\bibnamefont {Lejay}}, \bibinfo {author} {\bibfnamefont
  {S.}~\bibnamefont {Petit}}, \ and\ \bibinfo {author} {\bibfnamefont
  {R.}~\bibnamefont {Ballou}},\ }\href {\doibase 10.1103/PhysRevB.93.224401}
  {\bibfield  {journal} {\bibinfo  {journal} {Phys. Rev. B}\ }\textbf {\bibinfo
  {volume} {93}},\ \bibinfo {pages} {224401} (\bibinfo {year} {2016})},\
  \Eprint {http://arxiv.org/abs/1511.07486} {arXiv:1511.07486} \BibitemShut
  {NoStop}%
\bibitem [{\citenamefont {Michette}\ and\ \citenamefont
  {Pfauntsch}(2000)}]{Michette2000}%
  \BibitemOpen
  \bibfield  {author} {\bibinfo {author} {\bibfnamefont {A.~G.}\ \bibnamefont
  {Michette}}\ and\ \bibinfo {author} {\bibfnamefont {S.~J.}\ \bibnamefont
  {Pfauntsch}},\ }\href {\doibase 10.1088/0022-3727/33/10/308} {\bibfield
  {journal} {\bibinfo  {journal} {J. Phys. D. Appl. Phys.}\ }\textbf {\bibinfo
  {volume} {33}},\ \bibinfo {pages} {1186} (\bibinfo {year}
  {2000})}\BibitemShut {NoStop}%
\bibitem [{\citenamefont {Gretarsson}\ \emph {et~al.}(2013)\citenamefont
  {Gretarsson}, \citenamefont {Clancy}, \citenamefont {Liu}, \citenamefont
  {Hill}, \citenamefont {Bozin}, \citenamefont {Singh}, \citenamefont {Manni},
  \citenamefont {Gegenwart}, \citenamefont {Kim}, \citenamefont {Said},
  \citenamefont {Casa}, \citenamefont {Gog}, \citenamefont {Upton},
  \citenamefont {Kim}, \citenamefont {Yu}, \citenamefont {Katukuri},
  \citenamefont {Hozoi}, \citenamefont {{Van Den Brink}},\ and\ \citenamefont
  {Kim}}]{Gretarsson2013}%
  \BibitemOpen
  \bibfield  {author} {\bibinfo {author} {\bibfnamefont {H.}~\bibnamefont
  {Gretarsson}}, \bibinfo {author} {\bibfnamefont {J.~P.}\ \bibnamefont
  {Clancy}}, \bibinfo {author} {\bibfnamefont {X.}~\bibnamefont {Liu}},
  \bibinfo {author} {\bibfnamefont {J.~P.}\ \bibnamefont {Hill}}, \bibinfo
  {author} {\bibfnamefont {E.}~\bibnamefont {Bozin}}, \bibinfo {author}
  {\bibfnamefont {Y.}~\bibnamefont {Singh}}, \bibinfo {author} {\bibfnamefont
  {S.}~\bibnamefont {Manni}}, \bibinfo {author} {\bibfnamefont
  {P.}~\bibnamefont {Gegenwart}}, \bibinfo {author} {\bibfnamefont
  {J.}~\bibnamefont {Kim}}, \bibinfo {author} {\bibfnamefont {A.~H.}\
  \bibnamefont {Said}}, \bibinfo {author} {\bibfnamefont {D.}~\bibnamefont
  {Casa}}, \bibinfo {author} {\bibfnamefont {T.}~\bibnamefont {Gog}}, \bibinfo
  {author} {\bibfnamefont {M.~H.}\ \bibnamefont {Upton}}, \bibinfo {author}
  {\bibfnamefont {H.~S.}\ \bibnamefont {Kim}}, \bibinfo {author} {\bibfnamefont
  {J.}~\bibnamefont {Yu}}, \bibinfo {author} {\bibfnamefont {V.~M.}\
  \bibnamefont {Katukuri}}, \bibinfo {author} {\bibfnamefont {L.}~\bibnamefont
  {Hozoi}}, \bibinfo {author} {\bibfnamefont {J.}~\bibnamefont {{Van Den
  Brink}}}, \ and\ \bibinfo {author} {\bibfnamefont {Y.~J.}\ \bibnamefont
  {Kim}},\ }\href {\doibase 10.1103/PhysRevLett.110.076402} {\bibfield
  {journal} {\bibinfo  {journal} {Phys. Rev. Lett.}\ }\textbf {\bibinfo
  {volume} {110}},\ \bibinfo {pages} {1} (\bibinfo {year} {2013})},\ \Eprint
  {http://arxiv.org/abs/arXiv:1209.5424v1} {arXiv:arXiv:1209.5424v1}
  \BibitemShut {NoStop}%
\bibitem [{\citenamefont {Kim}\ \emph {et~al.}(2014{\natexlab{b}})\citenamefont
  {Kim}, \citenamefont {Daghofer}, \citenamefont {Said}, \citenamefont {Gog},
  \citenamefont {van~den Brink}, \citenamefont {Khaliullin},\ and\
  \citenamefont {Kim}}]{Kim2014}%
  \BibitemOpen
  \bibfield  {author} {\bibinfo {author} {\bibfnamefont {J.}~\bibnamefont
  {Kim}}, \bibinfo {author} {\bibfnamefont {M.}~\bibnamefont {Daghofer}},
  \bibinfo {author} {\bibfnamefont {a.~H.}\ \bibnamefont {Said}}, \bibinfo
  {author} {\bibfnamefont {T.}~\bibnamefont {Gog}}, \bibinfo {author}
  {\bibfnamefont {J.}~\bibnamefont {van~den Brink}}, \bibinfo {author}
  {\bibfnamefont {G.}~\bibnamefont {Khaliullin}}, \ and\ \bibinfo {author}
  {\bibfnamefont {B.~J.}\ \bibnamefont {Kim}},\ }\href {\doibase
  10.1038/ncomms5453} {\bibfield  {journal} {\bibinfo  {journal} {Nat.
  Commun.}\ }\textbf {\bibinfo {volume} {5}},\ \bibinfo {pages} {4453}
  (\bibinfo {year} {2014}{\natexlab{b}})},\ \Eprint
  {http://arxiv.org/abs/arXiv:1408.0804v1} {arXiv:arXiv:1408.0804v1}
  \BibitemShut {NoStop}%
\bibitem [{\citenamefont {{Moretti Sala}}\ \emph
  {et~al.}(2014{\natexlab{b}})\citenamefont {{Moretti Sala}}, \citenamefont
  {Boseggia}, \citenamefont {McMorrow},\ and\ \citenamefont
  {Monaco}}]{MorettiSala2014a}%
  \BibitemOpen
  \bibfield  {author} {\bibinfo {author} {\bibfnamefont {M.}~\bibnamefont
  {{Moretti Sala}}}, \bibinfo {author} {\bibfnamefont {S.}~\bibnamefont
  {Boseggia}}, \bibinfo {author} {\bibfnamefont {D.~F.}\ \bibnamefont
  {McMorrow}}, \ and\ \bibinfo {author} {\bibfnamefont {G.}~\bibnamefont
  {Monaco}},\ }\href {\doibase 10.1103/PhysRevLett.112.026403} {\bibfield
  {journal} {\bibinfo  {journal} {Phys. Rev. Lett.}\ }\textbf {\bibinfo
  {volume} {112}},\ \bibinfo {pages} {026403} (\bibinfo {year}
  {2014}{\natexlab{b}})}\BibitemShut {NoStop}%
\bibitem [{\citenamefont {Ament}\ \emph {et~al.}(2011)\citenamefont {Ament},
  \citenamefont {Khaliullin},\ and\ \citenamefont {{Van Den
  Brink}}}]{Ament2011}%
  \BibitemOpen
  \bibfield  {author} {\bibinfo {author} {\bibfnamefont {L.~J.~P.}\
  \bibnamefont {Ament}}, \bibinfo {author} {\bibfnamefont {G.}~\bibnamefont
  {Khaliullin}}, \ and\ \bibinfo {author} {\bibfnamefont {J.}~\bibnamefont
  {{Van Den Brink}}},\ }\href {\doibase 10.1103/PhysRevB.84.020403} {\bibfield
  {journal} {\bibinfo  {journal} {Phys. Rev. B - Condens. Matter Mater. Phys.}\
  }\textbf {\bibinfo {volume} {84}},\ \bibinfo {pages} {2} (\bibinfo {year}
  {2011})}\BibitemShut {NoStop}%
\bibitem [{\citenamefont {Liu}\ \emph {et~al.}(2012)\citenamefont {Liu},
  \citenamefont {Katukuri}, \citenamefont {Hozoi}, \citenamefont {Yin},
  \citenamefont {Dean}, \citenamefont {Upton}, \citenamefont {Kim},
  \citenamefont {Casa}, \citenamefont {Said}, \citenamefont {Gog},
  \citenamefont {Qi}, \citenamefont {Cao}, \citenamefont {Tsvelik},
  \citenamefont {{Van Den Brink}},\ and\ \citenamefont {Hill}}]{Liu2012}%
  \BibitemOpen
  \bibfield  {author} {\bibinfo {author} {\bibfnamefont {X.}~\bibnamefont
  {Liu}}, \bibinfo {author} {\bibfnamefont {V.~M.}\ \bibnamefont {Katukuri}},
  \bibinfo {author} {\bibfnamefont {L.}~\bibnamefont {Hozoi}}, \bibinfo
  {author} {\bibfnamefont {W.~G.}\ \bibnamefont {Yin}}, \bibinfo {author}
  {\bibfnamefont {M.~P.~M.}\ \bibnamefont {Dean}}, \bibinfo {author}
  {\bibfnamefont {M.~H.}\ \bibnamefont {Upton}}, \bibinfo {author}
  {\bibfnamefont {J.}~\bibnamefont {Kim}}, \bibinfo {author} {\bibfnamefont
  {D.}~\bibnamefont {Casa}}, \bibinfo {author} {\bibfnamefont {A.}~\bibnamefont
  {Said}}, \bibinfo {author} {\bibfnamefont {T.}~\bibnamefont {Gog}}, \bibinfo
  {author} {\bibfnamefont {T.~F.}\ \bibnamefont {Qi}}, \bibinfo {author}
  {\bibfnamefont {G.}~\bibnamefont {Cao}}, \bibinfo {author} {\bibfnamefont
  {A.~M.}\ \bibnamefont {Tsvelik}}, \bibinfo {author} {\bibfnamefont
  {J.}~\bibnamefont {{Van Den Brink}}}, \ and\ \bibinfo {author} {\bibfnamefont
  {J.~P.}\ \bibnamefont {Hill}},\ }\href {\doibase
  10.1103/PhysRevLett.109.157401} {\bibfield  {journal} {\bibinfo  {journal}
  {Phys. Rev. Lett.}\ }\textbf {\bibinfo {volume} {109}},\ \bibinfo {pages} {2}
  (\bibinfo {year} {2012})},\ \Eprint {http://arxiv.org/abs/1209.5294}
  {arXiv:1209.5294} \BibitemShut {NoStop}%
\bibitem [{\citenamefont {Boseggia}\ \emph {et~al.}(2013)\citenamefont
  {Boseggia}, \citenamefont {Walker}, \citenamefont {Vale}, \citenamefont
  {Springell}, \citenamefont {Feng}, \citenamefont {Perry}, \citenamefont
  {{Moretti Sala}}, \citenamefont {R{\o}nnow}, \citenamefont {Collins},\ and\
  \citenamefont {McMorrow}}]{Boseggia2013}%
  \BibitemOpen
  \bibfield  {author} {\bibinfo {author} {\bibfnamefont {S.}~\bibnamefont
  {Boseggia}}, \bibinfo {author} {\bibfnamefont {H.~C.}\ \bibnamefont
  {Walker}}, \bibinfo {author} {\bibfnamefont {J.}~\bibnamefont {Vale}},
  \bibinfo {author} {\bibfnamefont {R.}~\bibnamefont {Springell}}, \bibinfo
  {author} {\bibfnamefont {Z.}~\bibnamefont {Feng}}, \bibinfo {author}
  {\bibfnamefont {R.~S.}\ \bibnamefont {Perry}}, \bibinfo {author}
  {\bibfnamefont {M.}~\bibnamefont {{Moretti Sala}}}, \bibinfo {author}
  {\bibfnamefont {H.~M.}\ \bibnamefont {R{\o}nnow}}, \bibinfo {author}
  {\bibfnamefont {S.~P.}\ \bibnamefont {Collins}}, \ and\ \bibinfo {author}
  {\bibfnamefont {D.~F.}\ \bibnamefont {McMorrow}},\ }\href {\doibase
  10.1088/0953-8984/25/42/422202} {\bibfield  {journal} {\bibinfo  {journal}
  {J. Phys. Condens. Matter}\ }\textbf {\bibinfo {volume} {25}},\ \bibinfo
  {pages} {422202} (\bibinfo {year} {2013})},\ \Eprint
  {http://arxiv.org/abs/arXiv:1308.6460v2} {arXiv:arXiv:1308.6460v2}
  \BibitemShut {NoStop}%
\bibitem [{\citenamefont {Perkins}\ \emph {et~al.}(2014)\citenamefont
  {Perkins}, \citenamefont {Sizyuk},\ and\ \citenamefont
  {W{\"{o}}lfle}}]{Perkins2014}%
  \BibitemOpen
  \bibfield  {author} {\bibinfo {author} {\bibfnamefont {N.~B.}\ \bibnamefont
  {Perkins}}, \bibinfo {author} {\bibfnamefont {Y.}~\bibnamefont {Sizyuk}}, \
  and\ \bibinfo {author} {\bibfnamefont {P.}~\bibnamefont {W{\"{o}}lfle}},\
  }\href {\doibase 10.1103/PhysRevB.89.035143} {\bibfield  {journal} {\bibinfo
  {journal} {Phys. Rev. B - Condens. Matter Mater. Phys.}\ }\textbf {\bibinfo
  {volume} {89}},\ \bibinfo {pages} {035143} (\bibinfo {year} {2014})},\
  \Eprint {http://arxiv.org/abs/1311.0852} {arXiv:1311.0852} \BibitemShut
  {NoStop}%
\bibitem [{\citenamefont {{Moretti Sala}}\ \emph
  {et~al.}(2014{\natexlab{c}})\citenamefont {{Moretti Sala}}, \citenamefont
  {Rossi}, \citenamefont {Al-Zein}, \citenamefont {Boseggia}, \citenamefont
  {Hunter}, \citenamefont {Perry}, \citenamefont {Prabhakaran}, \citenamefont
  {Boothroyd}, \citenamefont {Brookes}, \citenamefont {McMorrow}, \citenamefont
  {Monaco},\ and\ \citenamefont {Krisch}}]{MorettiSala2014c}%
  \BibitemOpen
  \bibfield  {author} {\bibinfo {author} {\bibfnamefont {M.}~\bibnamefont
  {{Moretti Sala}}}, \bibinfo {author} {\bibfnamefont {M.}~\bibnamefont
  {Rossi}}, \bibinfo {author} {\bibfnamefont {A.}~\bibnamefont {Al-Zein}},
  \bibinfo {author} {\bibfnamefont {S.}~\bibnamefont {Boseggia}}, \bibinfo
  {author} {\bibfnamefont {E.~C.}\ \bibnamefont {Hunter}}, \bibinfo {author}
  {\bibfnamefont {R.~S.}\ \bibnamefont {Perry}}, \bibinfo {author}
  {\bibfnamefont {D.}~\bibnamefont {Prabhakaran}}, \bibinfo {author}
  {\bibfnamefont {A.~T.}\ \bibnamefont {Boothroyd}}, \bibinfo {author}
  {\bibfnamefont {N.~B.}\ \bibnamefont {Brookes}}, \bibinfo {author}
  {\bibfnamefont {D.~F.}\ \bibnamefont {McMorrow}}, \bibinfo {author}
  {\bibfnamefont {G.}~\bibnamefont {Monaco}}, \ and\ \bibinfo {author}
  {\bibfnamefont {M.}~\bibnamefont {Krisch}},\ }\href {\doibase
  10.1103/PhysRevB.90.085126} {\bibfield  {journal} {\bibinfo  {journal} {Phys.
  Rev. B}\ }\textbf {\bibinfo {volume} {90}},\ \bibinfo {pages} {085126}
  (\bibinfo {year} {2014}{\natexlab{c}})}\BibitemShut {NoStop}%
\bibitem [{\citenamefont {{Moretti Sala}}\ \emph
  {et~al.}(2014{\natexlab{d}})\citenamefont {{Moretti Sala}}, \citenamefont
  {Rossi}, \citenamefont {Boseggia}, \citenamefont {Akimitsu}, \citenamefont
  {Brookes}, \citenamefont {Isobe}, \citenamefont {Minola}, \citenamefont
  {Okabe}, \citenamefont {R{\o}nnow}, \citenamefont {Simonelli}, \citenamefont
  {McMorrow},\ and\ \citenamefont {Monaco}}]{MorettiSala2014d}%
  \BibitemOpen
  \bibfield  {author} {\bibinfo {author} {\bibfnamefont {M.}~\bibnamefont
  {{Moretti Sala}}}, \bibinfo {author} {\bibfnamefont {M.}~\bibnamefont
  {Rossi}}, \bibinfo {author} {\bibfnamefont {S.}~\bibnamefont {Boseggia}},
  \bibinfo {author} {\bibfnamefont {J.}~\bibnamefont {Akimitsu}}, \bibinfo
  {author} {\bibfnamefont {N.~B.}\ \bibnamefont {Brookes}}, \bibinfo {author}
  {\bibfnamefont {M.}~\bibnamefont {Isobe}}, \bibinfo {author} {\bibfnamefont
  {M.}~\bibnamefont {Minola}}, \bibinfo {author} {\bibfnamefont
  {H.}~\bibnamefont {Okabe}}, \bibinfo {author} {\bibfnamefont {H.~M.}\
  \bibnamefont {R{\o}nnow}}, \bibinfo {author} {\bibfnamefont {L.}~\bibnamefont
  {Simonelli}}, \bibinfo {author} {\bibfnamefont {D.~F.}\ \bibnamefont
  {McMorrow}}, \ and\ \bibinfo {author} {\bibfnamefont {G.}~\bibnamefont
  {Monaco}},\ }\href {\doibase 10.1103/PhysRevB.89.121101} {\bibfield
  {journal} {\bibinfo  {journal} {Phys. Rev. B}\ }\textbf {\bibinfo {volume}
  {89}},\ \bibinfo {pages} {121101} (\bibinfo {year}
  {2014}{\natexlab{d}})}\BibitemShut {NoStop}%
\bibitem [{\citenamefont {{Di Matteo}}\ and\ \citenamefont
  {Norman}(2016)}]{DiMatteo2016}%
  \BibitemOpen
  \bibfield  {author} {\bibinfo {author} {\bibfnamefont {S.}~\bibnamefont {{Di
  Matteo}}}\ and\ \bibinfo {author} {\bibfnamefont {M.~R.}\ \bibnamefont
  {Norman}},\ }\href {\doibase 10.1103/PhysRevB.94.075148} {\bibfield
  {journal} {\bibinfo  {journal} {Phys. Rev. B}\ }\textbf {\bibinfo {volume}
  {94}},\ \bibinfo {pages} {1} (\bibinfo {year} {2016})},\ \Eprint
  {http://arxiv.org/abs/1603.04311} {arXiv:1603.04311} \BibitemShut {NoStop}%
\bibitem [{\citenamefont {Jørgensen}(1963)}]{Jorgensen1963}%
  \BibitemOpen
  \bibfield  {author} {\bibinfo {author} {\bibfnamefont {C.~K.}\ \bibnamefont
  {Jørgensen}},\ }in\ \href {http://doi.wiley.com/10.1002/9780470143513.ch2}
  {\emph {\bibinfo {booktitle} {Advances in Chemical Physics}}},\ Vol.~\bibinfo
  {volume} {5},\ \bibinfo {editor} {edited by\ \bibinfo {editor} {\bibfnamefont
  {I.}~\bibnamefont {Prigogine}}}\ (\bibinfo  {publisher} {John Wiley \& Sons,
  Inc.},\ \bibinfo {year} {1963})\ pp.\ \bibinfo {pages} {33--146},\ \bibinfo
  {note} {{DOI}: 10.1002/9780470143513.ch2}\BibitemShut {NoStop}%
\bibitem [{\citenamefont {Agrestini}\ \emph {et~al.}(2017)\citenamefont
  {Agrestini}, \citenamefont {Kuo}, \citenamefont {Moretti~Sala}, \citenamefont
  {Hu}, \citenamefont {Kasinathan}, \citenamefont {Ko}, \citenamefont
  {Glatzel}, \citenamefont {Rossi}, \citenamefont {Cafun}, \citenamefont
  {Kvashnina}, \citenamefont {Matsumoto}, \citenamefont {Takayama},
  \citenamefont {Takagi}, \citenamefont {Tjeng},\ and\ \citenamefont
  {Haverkort}}]{Agrestini2017}%
  \BibitemOpen
  \bibfield  {author} {\bibinfo {author} {\bibfnamefont {S.}~\bibnamefont
  {Agrestini}}, \bibinfo {author} {\bibfnamefont {C.-Y.}\ \bibnamefont {Kuo}},
  \bibinfo {author} {\bibfnamefont {M.}~\bibnamefont {Moretti~Sala}}, \bibinfo
  {author} {\bibfnamefont {Z.}~\bibnamefont {Hu}}, \bibinfo {author}
  {\bibfnamefont {D.}~\bibnamefont {Kasinathan}}, \bibinfo {author}
  {\bibfnamefont {K.-T.}\ \bibnamefont {Ko}}, \bibinfo {author} {\bibfnamefont
  {P.}~\bibnamefont {Glatzel}}, \bibinfo {author} {\bibfnamefont
  {M.}~\bibnamefont {Rossi}}, \bibinfo {author} {\bibfnamefont {J.-D.}\
  \bibnamefont {Cafun}}, \bibinfo {author} {\bibfnamefont {K.~O.}\ \bibnamefont
  {Kvashnina}}, \bibinfo {author} {\bibfnamefont {A.}~\bibnamefont
  {Matsumoto}}, \bibinfo {author} {\bibfnamefont {T.}~\bibnamefont {Takayama}},
  \bibinfo {author} {\bibfnamefont {H.}~\bibnamefont {Takagi}}, \bibinfo
  {author} {\bibfnamefont {L.~H.}\ \bibnamefont {Tjeng}}, \ and\ \bibinfo
  {author} {\bibfnamefont {M.~W.}\ \bibnamefont {Haverkort}},\ }\href {\doibase
  10.1103/PhysRevB.95.205123} {\bibfield  {journal} {\bibinfo  {journal} {Phys.
  Rev. B}\ }\textbf {\bibinfo {volume} {95}},\ \bibinfo {pages} {205123}
  (\bibinfo {year} {2017})}\BibitemShut {NoStop}%
\bibitem [{\citenamefont {Vaknin}\ \emph {et~al.}(1987)\citenamefont {Vaknin},
  \citenamefont {Sinha}, \citenamefont {Moncton}, \citenamefont {Johnston},
  \citenamefont {Newsam}, \citenamefont {Safinya},\ and\ \citenamefont
  {King}}]{Vaknin1987}%
  \BibitemOpen
  \bibfield  {author} {\bibinfo {author} {\bibfnamefont {D.}~\bibnamefont
  {Vaknin}}, \bibinfo {author} {\bibfnamefont {S.~K.}\ \bibnamefont {Sinha}},
  \bibinfo {author} {\bibfnamefont {D.~E.}\ \bibnamefont {Moncton}}, \bibinfo
  {author} {\bibfnamefont {D.~C.}\ \bibnamefont {Johnston}}, \bibinfo {author}
  {\bibfnamefont {J.~M.}\ \bibnamefont {Newsam}}, \bibinfo {author}
  {\bibfnamefont {C.~R.}\ \bibnamefont {Safinya}}, \ and\ \bibinfo {author}
  {\bibfnamefont {H.~E.}\ \bibnamefont {King}},\ }\href {\doibase
  10.1103/PhysRevLett.58.2802} {\bibfield  {journal} {\bibinfo  {journal}
  {Phys. Rev. Lett.}\ }\textbf {\bibinfo {volume} {58}},\ \bibinfo {pages}
  {2802} (\bibinfo {year} {1987})}\BibitemShut {NoStop}%
\bibitem [{\citenamefont {Chen}\ \emph {et~al.}(1991)\citenamefont {Chen},
  \citenamefont {Sette}, \citenamefont {Ma}, \citenamefont {Hybertsen},
  \citenamefont {Stechel}, \citenamefont {Foulkes}, \citenamefont {Schulter},
  \citenamefont {Cheong}, \citenamefont {Cooper}, \citenamefont {Rupp},
  \citenamefont {Batlogg}, \citenamefont {Soo}, \citenamefont {Ming},
  \citenamefont {Krol},\ and\ \citenamefont {Kao}}]{Chen1991}%
  \BibitemOpen
  \bibfield  {author} {\bibinfo {author} {\bibfnamefont {C.}~\bibnamefont
  {Chen}}, \bibinfo {author} {\bibfnamefont {F.}~\bibnamefont {Sette}},
  \bibinfo {author} {\bibfnamefont {Y.}~\bibnamefont {Ma}}, \bibinfo {author}
  {\bibfnamefont {M.}~\bibnamefont {Hybertsen}}, \bibinfo {author}
  {\bibfnamefont {E.}~\bibnamefont {Stechel}}, \bibinfo {author} {\bibfnamefont
  {W.}~\bibnamefont {Foulkes}}, \bibinfo {author} {\bibfnamefont
  {M.}~\bibnamefont {Schulter}}, \bibinfo {author} {\bibfnamefont
  {S.}~\bibnamefont {Cheong}}, \bibinfo {author} {\bibfnamefont
  {A.}~\bibnamefont {Cooper}}, \bibinfo {author} {\bibfnamefont
  {L.}~\bibnamefont {Rupp}}, \bibinfo {author} {\bibfnamefont {B.}~\bibnamefont
  {Batlogg}}, \bibinfo {author} {\bibfnamefont {Y.}~\bibnamefont {Soo}},
  \bibinfo {author} {\bibfnamefont {Z.}~\bibnamefont {Ming}}, \bibinfo {author}
  {\bibfnamefont {A.}~\bibnamefont {Krol}}, \ and\ \bibinfo {author}
  {\bibfnamefont {Y.}~\bibnamefont {Kao}},\ }\href {\doibase
  10.1103/PhysRevLett.66.104} {\bibfield  {journal} {\bibinfo  {journal} {Phys.
  Rev. Lett.}\ }\textbf {\bibinfo {volume} {66}},\ \bibinfo {pages} {104}
  (\bibinfo {year} {1991})}\BibitemShut {NoStop}%
\bibitem [{\citenamefont {Ito}\ and\ \citenamefont {Akimitsu}(1976)}]{Ito1976}%
  \BibitemOpen
  \bibfield  {author} {\bibinfo {author} {\bibfnamefont {Y.}~\bibnamefont
  {Ito}}\ and\ \bibinfo {author} {\bibfnamefont {J.}~\bibnamefont {Akimitsu}},\
  }\href {\doibase 10.1143/JPSJ.40.1333} {\bibfield  {journal} {\bibinfo
  {journal} {Journal of the Physical Society of Japan}\ }\textbf {\bibinfo
  {volume} {40}},\ \bibinfo {pages} {1333} (\bibinfo {year} {1976})},\ \Eprint
  {http://arxiv.org/abs/http://dx.doi.org/10.1143/JPSJ.40.1333}
  {http://dx.doi.org/10.1143/JPSJ.40.1333} \BibitemShut {NoStop}%
\bibitem [{\citenamefont {Kugel'}\ and\ \citenamefont
  {Khomskii}(1982)}]{Kugel1982}%
  \BibitemOpen
  \bibfield  {author} {\bibinfo {author} {\bibfnamefont {K.~I.}\ \bibnamefont
  {Kugel'}}\ and\ \bibinfo {author} {\bibfnamefont {D.~I.}\ \bibnamefont
  {Khomskii}},\ }\href {http://stacks.iop.org/0038-5670/25/i=4/a=R03}
  {\bibfield  {journal} {\bibinfo  {journal} {Soviet Physics Uspekhi}\ }\textbf
  {\bibinfo {volume} {25}},\ \bibinfo {pages} {231} (\bibinfo {year}
  {1982})}\BibitemShut {NoStop}%
\bibitem [{\citenamefont {Plotnikova}\ \emph {et~al.}(2016)\citenamefont
  {Plotnikova}, \citenamefont {Daghofer}, \citenamefont {van~den Brink},\ and\
  \citenamefont {Wohlfeld}}]{Plotnikova2016}%
  \BibitemOpen
  \bibfield  {author} {\bibinfo {author} {\bibfnamefont {E.~M.}\ \bibnamefont
  {Plotnikova}}, \bibinfo {author} {\bibfnamefont {M.}~\bibnamefont
  {Daghofer}}, \bibinfo {author} {\bibfnamefont {J.}~\bibnamefont {van~den
  Brink}}, \ and\ \bibinfo {author} {\bibfnamefont {K.}~\bibnamefont
  {Wohlfeld}},\ }\href {\doibase 10.1103/PhysRevLett.116.106401} {\bibfield
  {journal} {\bibinfo  {journal} {Phys. Rev. Lett.}\ }\textbf {\bibinfo
  {volume} {116}},\ \bibinfo {pages} {106401} (\bibinfo {year} {2016})},\
  \Eprint {http://arxiv.org/abs/1601.07069} {arXiv:1601.07069} \BibitemShut
  {NoStop}%
\end{thebibliography}%

\end{document}